\documentclass[showpacs]{revtex4}
\usepackage{graphicx}
\newcommand{\et}{{\it et al.}\ }
\newcommand{\bulk}{{\rm bulk}}
\newcommand{\hs}{{\rm hs}}
\newcommand{\wf}{{\rm wf}}
\renewcommand{\lg}{{\rm lg}}

\begin{document}
\title{Wall-Fluid and Liquid-Gas Interfaces of Model Colloid-Polymer Mixtures by Simulation and Theory}

\author{Andrea Fortini}
\email{a.fortini@phys.uu.nl}
\author{Marjolein Dijkstra}
\affiliation{Soft Condensed Matter, Debye Institute, Utrecht University,
   Princetonplein 5, 3584 CC Utrecht, The Netherlands.} 
\author{Matthias Schmidt\footnote{Present address:
 Institut f\"ur Theoretische Physik II,
 Heinrich-Heine-Universit\"at D\"usseldorf,
 Universit\"atsstra\ss e 1, D-40225 D\"usseldorf, Germany.}}
\affiliation{Soft Condensed Matter, Debye Institute, Utrecht University,
   Princetonplein 5, 3584 CC Utrecht, The Netherlands.}
\author{Paul P.~F.  Wessels}
\affiliation{Institut f\"ur Theoretische Physik II,
 Heinrich-Heine-Universit\"at D\"usseldorf,
 Universit\"atsstra\ss e 1, D-40225 D\"usseldorf, Germany.}


\begin{abstract}
We perform a study of the interfacial properties of a model suspension
of hard sphere colloids with diameter $\sigma_c$ and non-adsorbing
ideal polymer coils with diameter $\sigma_p$. For the mixture in
contact with a planar hard wall, we obtain from simulations the
wall-fluid interfacial free energy, $\gamma_{\rm wf}$, for size ratios
$q=\sigma_p/\sigma_c=0.6$ and 1, using thermodynamic integration, and
study the (excess) adsorption of colloids, $\Gamma_c$, and of
polymers, $\Gamma_p$, at the hard wall. The interfacial tension of the
free liquid-gas interface, $\gamma_{\lg}$, is obtained following three
different routes in simulations: i) from
studying the system size dependence of the interfacial width according
to the predictions of capillary wave theory, ii) from the probability
distribution of the colloid density at coexistence in the grand
canonical ensemble, and iii) for statepoints where the colloidal
liquid wets the wall completely, from Young's equation relating
$\gamma_{\lg}$ to the difference of wall-liquid and wall-gas
interfacial tensions, $\gamma_{\rm wl}-\gamma_{\rm wg}$. In addition,
we calculate $\gamma_{\rm wf}, \Gamma_c$, and $\Gamma_p$ using density
functional theory and a scaled particle theory based on free volume
theory.  Good agreement is found between the simulation results and
those from density functional theory, while the results from scaled
particle theory quantitatively deviate but reproduce some essential
features. Simulation results for $\gamma_{\lg}$ obtained from the
three different routes are all in good agreement.  Density functional
theory predicts $\gamma_{\lg}$ with good accuracy for high polymer
reservoir packing fractions, but yields deviations from the simulation
results close to the critical point.
\end{abstract}
\maketitle

\section{Introduction}
Mixtures of sterically-stabilized colloids and non-adsorbing polymers
are widely studied complex fluids \cite{WCKP02,RT03,JMB03}. Provided
the size and the number of polymers is sufficiently high, such
mixtures can phase-separate into a colloidal gas phase that is poor in
colloids and rich in polymers, and a colloidal liquid phase that is
rich in colloids and poor in polymers. The mechanism behind this
demixing transition is of entropic origin and is due to the so-called
depletion effect: An effective attraction between the colloids is
generated due to an unbalanced osmotic pressure arising from exclusion
of polymer coils in the depletion zones between the colloids
\cite{SAFO54,AV76}.  Since the associated relevant time and length
scales are much larger than in atomic and molecular systems, direct
experimental observations using advanced microscopy techniques enable
the study of many interesting physical phenomena, e.g.\ thermal
capillary waves at fluid interfaces and droplet coalescence were
observed recently in real space and real time using confocal
microscopy \cite{DGALA04b}.  Other recent examples are the direct
measurement of the contact angle of the colloidal gas-liquid interface
and different substrates \cite{WKW03-a}, complete wetting of a
substrate \cite{DGALA03,WKW03}, a wetting transition from complete to
partial wetting \cite{WKW03}, and capillary condensation in confining
geometry \cite{DGALA04}. Hence it is fair to say that these mixtures
serve as excellent model systems to investigate fundamental issues in
statistical physics.  A particularly simple model for colloid-polymer
mixtures was proposed by Vrij \cite{AV76}, and is often referred to as
the Asakura-Oosawa-Vrij (AOV) model. A good historical introduction to
this model together with many recent results can be found in the paper
by Brader \et \cite{JMB03}. The bulk phase behavior as well as
inhomogeneous properties of the AOV model were studied with both
theory \cite{HNWL+92,JMB03,DGALA+04,PPFW04,PPFW04a} and computer simulation \cite{EJM94,MDJMBRE99,MDRvR02,PBAAL02-a,MSAF03,MSAF04,RLCV03,RLCV04,RLCV04b}.
Recently, attention was paid to more realistic model interactions
\cite{JDAJ+01,PBAAL02-a,PPFW04a} for colloid-polymer
mixtures. However, most of the essential physics of real
colloid-polymer mixtures is indeed captured by the AOV model.

In this work we study the wall-fluid and liquid-gas interfacial
tensions of the AOV colloid-polymer mixture using Monte Carlo
simulations and check the predictions of density functional theory
(DFT) \cite{MSHL+00,PPFW04a} based on an extension of the Rosenfeld
functional \cite{YR89}, and of a scaled particle theory (SPT)
\cite{PPFW04a} based on the free volume theory \cite{HNWL+92}. The
wall-fluid interfacial tension of a hard-sphere fluid in contact with
a planar hard wall was recently calculated by Heni and L\"owen using a
thermodynamic integration procedure along a path that corresponds to
the growth of a wall in a bulk system \cite{MHHL99}.  Here, we propose
a thermodynamic integration approach similar in spirit, to determine
the wall-fluid interfacial free energy of the AOV colloid-polymer
mixture in contact with a planar hard wall.

Different routes exist to determine the liquid-gas interfacial
tension. The pressure tensor method \cite{JRH84,HHWP04} is
particularly suitable for Molecular Dynamics simulation studies. The
probability distribution method \cite{KB82} was used in Monte-Carlo
simulations \cite{JJP00,MMLGM03} and was recently applied to calculate
the liquid-gas interfacial tension of colloid-polymer mixtures
\cite{RLCV03,RLCV04,RLCV04b,RLCV05}.  The authors showed that the
interfacial width depends on system size and they verified the
predictions from capillary wave theory on system size dependence
\cite{RLCV04}.  Sides \et and Lacasse \et found good agreement between
the interfacial tension obtained from the pressure tensor method and
from the predictions of capillary wave theory provided that the
density profile is fit to an error function for Lennard-Jones
particles and polymer blends \cite{sides,lacasse}.  In this work, we
use both the capillary wave theory, as proposed in Refs.\
\cite{sides,lacasse}, as well as the probability distribution method
to calculate the liquid-gas interfacial tension. In addition, we
employ Young's equation in the complete wetting regime to obtain an
estimate for the tension of the liquid-gas interface from the
difference in wall tensions, and compare our results with DFT
calculations.

The paper is organized as follows. In section \ref{S:model} we briefly
review the AOV model.  In section \ref{S:tension} an overview of the
thermodynamics of inhomogeneous systems is given. In section
\ref{S:simwf} we lay out the simulation methods used in the
determination of the wall-fluid tension.  Section \ref{S:simads}
details the simulation method to obtain the adsorption at a hard wall.
Section \ref{S:spt} presents the corresponding derivation using scaled
particle theory.  In section \ref{S:capwave} we review the capillary
wave theory for the width of a liquid-gas interface. Sections
\ref{S:probabilityDistribution} and \ref{S:young} are devoted to
obtaining the liquid-gas interfacial tension from the probability
distribution of the colloid density at coexistence and from Young's
equation, respectively. Section \ref{S:dft} gives a brief overview of
the DFT. In section \ref{S:res} we discuss our results for the
wall-fluid tension (\ref{S:reswf}), adsorption (\ref{S:resads}), and
liquid-gas interfacial tension (\ref{S:reslg}).  Concluding remarks
are given in section \ref{S:end}.

\section{Model}
\label{S:model} 
We consider a mixture of sterically-stabilized colloidal spheres
(species $c$) and non-adsorbing polymer coils (species $p$) immersed
in a solvent. The interaction between two sterically stabilized
colloids resembles closely that of hard spheres, while a dilute
solution of polymer coils in a theta-solvent can be treated as an
ideal gas as the polymer coils are interpenetrable and
noninteracting. The polymer coils are assumed to be excluded from the
colloids to a center-of-mass distance of $(\sigma_c+\sigma_p)/2$,
where $\sigma_c$ is the diameter of the colloids, and $\sigma_p=2 R_g$
is the diameter of the polymer coils, with $R_g$ being the radius of
gyration of the polymers. A simple idealized model for such a mixture
is the so-called Asakura-Oosawa-Vrij (AOV) model \cite{SAFO54,AV76}
defined through pair potentials, that between colloids being
\begin{equation}
v_{cc}(R_{ij})= \left \{ \begin{array}{ll}
\infty & \textrm{ if $R_{ij} < \sigma_c$ } \\
0 & \textrm{ otherwise,}
\end{array} \right .
\label{E:vcc}
\end{equation}
where $R_{ij}=|\vec{R}_i-\vec{R}_j|$ is the center-of-mass
distance between two colloidal particles, with $\vec{R}_i$ ($\vec{R}_j$) the
center-of-mass coordinate of colloid $i$ ($j$). The polymers are
described as noninteracting particles with
\begin{equation}
v_{pp}(r_{ij})= 0  ,
\end{equation}
where $r_{ij}=|\vec{r}_i-\vec{r}_j|$ is the distance between two
polymers, with $\vec{r}_i$ ($\vec{r}_j$) the center-of-mass coordinate of
polymer $i$ ($j$). 
The colloid-polymer interaction is
\begin{equation}
v_{cp}(|\vec{R}_{i}-\vec{r}_{j}|)= \left \{ \begin{array}{ll}
\infty & \textrm{ if $|\vec{R}_{i}-\vec{r}_{j}| < (\sigma_c+\sigma_p)/2$ } \\
0 & \textrm{ otherwise,}
\end{array} \right .
\end{equation}
where $|\vec{R}_i-\vec{r}_j|$ is the distance between colloid $i$ and
polymer $j$.  Our simulations are performed in a box with dimensions $
L\times L \times H $ with three-dimensional periodic boundary
conditions in the case of bulk simulations.  To create a wall-fluid or
a liquid-gas interface, we perform simulations in a box with periodic
boundary condition solely in the $x$ and $y$ directions and two
impenetrable hard walls in the $z$ direction. The wall-particle
potential acting on particles of species $k=c,p$ reads
\begin{equation}
v_{wk}(z_{k,i})= \left \{ \begin{array}{ll}
0 & \textrm{ if $  \sigma_k/2  < z_{k,i}  < H_k-\sigma_k/2 $ } \\
\infty & \textrm{ otherwise}
\end{array} \right . ,
\label{E:hw}
\end{equation}
where $z_{k,i}$ is the $z$-coordinate of particle $i$ of species $k$,
and $H_k$ is the separation distance between the two walls for species
$k$. For the simulations of hard wall properties, we use
$H_c=H_p\equiv H$, corresponding to two planar hard walls.  For the
simulations of the liquid-gas interface we use a box with periodic
boundary conditions in the $x$ and $y$ directions and delimited in the
$z$ direction by one impenetrable and one semi-permeable wall; this
asymmetric slit is defined by the wall-particle potential (\ref{E:hw})
with $H_c=H-2\sigma_c$ and $H_p=H$.  The impenetrable wall at $z=0$
favors the colloidal {\it liquid} phase, while the semi-permeable wall
at $z=H_c$ is impenetrable for the colloidal particles, but penetrable
for the polymers (which are free to overlap with this wall).  Hence,
there is no effective polymer-mediated wall-colloid attraction and the
semi-permeable wall favors the colloidal gas phase
\cite{MDRvR02,MSAF03,MSAF04,PPFW04a}.

The size ratio $q=\sigma_p/\sigma_c$ is a geometric parameter that
controls the range of the effective depletion interaction between the
colloids.  Packing fraction are denoted by $\eta_k=(\pi\sigma_k^3
N_k)/(6 L^2 H)$, and the number density is denoted by $\rho_k =
N_k/(L^2H)$ for species $k=c,p$.  We also employ, as an alternative
thermodynamic variable to $\eta_p$, the polymer reservoir packing
fraction
\begin{equation}
\eta_p^r =\frac{\pi}{6} \sigma_p^3 \rho_p^r =\frac{\pi}{6}
\sigma_p^3 z_p,
\label{pacres}
\end{equation}
where $z_p$ is the fugacity of the polymers, that constitute an ideal
gas with density $\rho_p^r$ in the reservoir.

\section{Methods}
\subsection{Overview of interfacial thermodynamics}
\label{S:tension} 
Generally, the interfacial tension in an inhomogeneous system is the
grand potential per unit area needed to create an interface in an
initially uniform bulk system at fixed chemical potential of colloids,
$\mu_c$, and polymers, $\mu_p$, and fixed volume $V$ and temperature
$T$.  The grand potential for a bulk mixture of colloids and polymers
reads
\begin{equation}
\Omega^\bulk( \mu_c, \mu_p, V, T)=-p(\mu_c,\mu_p,T) V,
\label{fgranpotbulk}
\end{equation}
where $p$ is the bulk pressure.  The system
in contact with an interface possesses the grand potential
\begin{equation}
\Omega( \mu_c, \mu_p,V, T,A)=-p(\mu_c,\mu_p,T) V  +  \gamma (\mu_c,\mu_p,T) A , 
\label{fgranpotsur}
\end{equation}
where $A$ is the area of the interface and $\gamma(\mu_c,\mu_p,T)$ is
the interfacial tension, which can hence be expressed as
\begin{equation}
 \gamma =\frac{\Omega( \mu_c, \mu_p, V, T,A)-
   \Omega^\bulk( \mu_c, \mu_p, V, T)}{A}.
\label{fgamma}
\end{equation}
Besides the liquid-gas interface, where $\gamma=\gamma_{\rm lg}$,
Eqs.\ (\ref{fgranpotbulk}), (\ref{fgranpotsur}), and (\ref{fgamma})
apply also for a fluid adsorbed between two parallel plates (walls),
where $\gamma=\gamma_{\rm wf}$, provided that the wall separation is
sufficiently large \cite{REUMBM87,MD97b}, and that the area $A$ is
equal to the total area of the two plates, $A=2 L^2$.  At fixed
chemical potentials the number of particles in the inhomogeneous
system, $N_c$ and $N_p$, of colloids and polymers, respectively, will
be in general different from those in the bulk, $N_c^\bulk$ and
$N_p^\bulk$. The excess number of colloids and polymers per unit area,
i.e. the adsorptions $\Gamma_c$ and $\Gamma_p$, respectively, are
defined as
\begin{eqnarray}
\Gamma_c(\mu_c, \mu_p, T)& =& \frac{ N_c- N_c^\bulk}{A} \label{Gammac} ,\\
\Gamma_p(\mu_c, \mu_p, T)& =& \frac{ N_p- N_p^\bulk}{A}.
\label{Gammap}
\end{eqnarray}
The grand potentials (\ref{fgranpotbulk}) and (\ref{fgranpotsur})
in differential form read \begin{eqnarray}
d\Omega^\bulk(\mu_c,\mu_p,V,T) &=&-N_c^\bulk d\mu_c-N_p^\bulk
d\mu_p -p dV -S^\bulk dT,
\label{domega},\\
d\Omega(\mu_c,\mu_p,V,T,A) &=&-N_c d\mu_c-N_p d\mu_p -p dV -SdT + \gamma dA.
\label{domegasurf}
\end{eqnarray} 
Using Eqs.\ (\ref{domega}) and (\ref{domegasurf}) and Eq.\
(\ref{fgamma}) in differential form, it is straightforward to show
\cite{JRBW82} that the adsorptions are related to the interfacial
tension through
\begin{equation}
\Gamma_c = - \left( \frac{\partial \gamma}{\partial \mu_c}
\right)_{\mu_p,T} 
\hspace{5mm}\mbox{and} \hspace{5mm} 
\Gamma_p = -
\left( \frac{\partial \gamma}{\partial \mu_p} \right)_{\mu_c,T}.
\label{E:ad}
\end{equation}

\subsection{Hard wall-fluid interfacial tension via thermodynamic integration}
\label{S:simwf}
To determine, from simulations, the wall-fluid tension
$\gamma_{\rm wf}$ of the AOV model we should apply equation (\ref{fgamma}),
as is manifest in the grand canonical ensemble, i.e. for constant
colloid and polymer fugacities.  However, in our simulation it is more
convenient to use the semi-grand canonical ensemble fixing the number
of colloids and the fugacity of the polymers. The reason is
twofold. First the interfacial tension as a function of the fugacity
of polymers can be directly compared to the DFT results of Ref.\
\cite{PPFW04a}. Second, fixing the number of colloids instead of their
fugacity allow us to efficiently study state points with high packing
fractions of colloids; generally grand ensemble simulations are
difficult to perform at high densities due to small particle insertion
probabilities.  
To compute the tension we have to recast Eq. (\ref{fgamma}) in a way that is
consistent with the semi-grand canonical ensemble.  The grand
potentials for the bulk and the inhomogeneous system are related to
the corresponding Helmholtz free energies via a Legendre
transformation,
\begin{eqnarray}
\Omega^\bulk(\mu_c,\mu_p,V,T) &=& F^\bulk(N_c^\bulk,N_p^\bulk,V,T) - \mu_c N_c^\bulk - \mu_p N_p^\bulk, \label{om1} \\
\Omega(\mu_c,\mu_p,V,T,A) &=& F(N_c,N_p,V,T,A) - \mu_c N_c - \mu_p N_p.\label{om2}
\end{eqnarray}
We substitute Eqs. (\ref{om1}) and (\ref{om2}) in Eq. (\ref{fgamma}) to obtain 
\begin{equation}
\gamma =\frac{F(N_c,N_p)-F^\bulk(N_c^\bulk,N_p^\bulk)}{A}-\mu_c
\Gamma_c -\mu_p\Gamma_p  ,
 \label{gammac}
\end{equation}
where we omitted the dependence on the variables $V,T,\mu_c,$ and
$\mu_p$ in the notation. Note that the tension is not only the
difference of the Helmholtz free energies, but additional terms,
$\mu_c \Gamma_c$ and $\mu_p \Gamma_p$, arise in Eq.\
(\ref{gammac}). One can further simplify by Taylor expanding
$F(N_c,N_p,V,T,A)$ around $N_c^\bulk$:
\begin{equation}
F(N_c,N_p,V,T,A)= F(N_c^\bulk,N_p,V,T,A)+\frac{\partial
F}{\partial N_c} (N_c-N_c^\bulk)+ \mathcal{O}( (N_c-N_c^\bulk)^2) .
\end{equation}
Keeping only the first order term, one can approximate the
interfacial tension as
\begin{equation}
\gamma \simeq
\frac{F(N_c^\bulk,N_p,V,T,A)-F^\bulk(N_c^\bulk,N_p^\bulk,V,T)}{A}
-\mu_p\Gamma_p. \label{E:gammahs}
\end{equation}
The same approximation was employed in Ref.\ \cite{MHHL99}
(using $N_p=0$ and $N_p^\bulk=0$) to calculate the interfacial free
energy of hard spheres in contact with a planar hard wall. 
To compute the wall tension, we need to perform {\em two}
free energy calculations, one for the bulk and one for the
inhomogeneous system.  As the free energy cannot be measured directly
in a Monte Carlo simulation, we use the thermodynamic integration
technique\cite{DFBS96} to relate the free energy of the system of interest to that
of a reference system 
\begin{equation}
F(N_c,N_p,V,T,A,\lambda=\lambda_{\rm max}) =
F_{id}(N_c,N_p,V,T,\lambda=0)+\int_{\lambda=0}^{\lambda_{\rm max}} d
\lambda \left\langle \frac{\partial F}{\partial \lambda}
\right\rangle_\lambda.
\label{thint}
\end{equation}
The reference system is chosen to be an ideal gas, so $F_{id}(N_c,N_p,V,T,\lambda=0)$ is the Helmholtz free energy of
$N_c$ ideal colloids and $N_p$ ideal polymers in a volume $V$ at
temperature $T$. 
We then introduce the suitable auxiliary Hamiltonian
\begin{equation} H_{\lambda}= \lambda \left
(\sum_{i<j}^{N_c} V_{cc}(R_{ij}) +
\sum_{i=1}^{N_c}\sum_{j=1}^{N_p} V_{cp}(|\vec{R}_i-\vec{r}_j|) +
\epsilon \sum_{i=1}^{N_c} V_{wc}(z_{c,i}) + \epsilon
\sum_{i=1}^{N_p} V_{wp}(z_{p,i}) \right ), \label{E:ham}
\end{equation}
where we approximate the hard-core potentials of the AOV model with
penetrable potentials: The colloid-colloid interaction reads
\begin{equation}
V_{cc}(R_{ij}) = \Theta(\sigma_c- R_{ij} ) , \label{E:shcc}
\end{equation}
where $\Theta(x)$ is the Heaviside step function.  
Likewise we define the interaction
potential between the colloids and the polymers as
\begin{equation}
V_{cp}(|\vec{R}_i-\vec{r}_j|) = \Theta \left(
\frac{(\sigma_c+\sigma_p)}{2}-|\vec{R}_i-\vec{r}_j| \right) .
\label{E:shcp}
\end{equation}
The interaction between the walls and the particles of species $k=c,p$ is
\begin{equation}
V_{wk}(z_{k,i}) = 
\Theta \left( \frac{\sigma_k}{2}-z_{k,i}  \right)+\Theta \left(
\frac{\sigma_k}{2}-(H_k-z_{k,i})  \right), \label{E:shws}
\end{equation}
where  $z_{k,i}$ is the $z$-coordinate of  particle $i$ of species
$k$. For $\lambda=0$, this system reduces to an ideal gas, while
for $\lambda\rightarrow \infty$, the system describes the AOV
model given by Eqs.\ (\ref{E:vcc})-(\ref{E:hw}) in bulk
($\epsilon=0$) or confined by two walls ($\epsilon=1$). 
The interaction potential is switched on
adiabatically using the coupling parameter $\lambda$. In principle,
our system of interest is described by the Hamiltonian (\ref{E:ham})
using $\lambda_{\rm max} \rightarrow \infty$, but also for
sufficiently high values of $\lambda_{\rm max}$ the system reduces to
our system of interest with hard-core potentials. Clearly,
$\lambda_{\rm max}$ should be sufficiently large to ensure that the
system is indeed behaving as the hard-core system of interest. On the
other hand, $\lambda_{\rm max}$ should not be too large, as this would
make the numerical integration less accurate. 
The integrand function of  Eq.\ (\ref{thint}) is then
\begin{equation}
\left \langle \frac{\partial F}{\partial \lambda} \right \rangle_{\lambda} =
 \left \langle \frac{\partial H_\lambda}{\partial \lambda} \right \rangle_{\lambda} =  \left \langle \frac{H_\lambda}{ \lambda} \right \rangle_{\lambda}. 
\end{equation}
The function $ \left \langle \frac{ H_\lambda}{ \lambda} \right \rangle_{\lambda} $ is computed counting the number of overlaps between  colloids, colloids and polymers and (for the inhomogeneous system only) particles and walls.  
The free energy  can then be obtained using Eq. (\ref{thint}). 
The integrals are evaluated using a 21-point Gauss-Kronrod formula, where 5000-15000 MC
cycles per particle are used for the sampling of each integration
point. 
The wall-fluid interfacial tension is then computed using equation
(\ref{E:gammahs}). In detail, we first perform a simulation of the
AOV model in bulk and a separate simulation of the AOV model confined
by two walls, given by the interactions (\ref{E:vcc})-(\ref{E:hw}),
both in the semi-grand canonical ensemble, i.e., we fix the number of
colloids $N_c^\bulk$, the chemical potential of the polymers $\mu_p$,
and the volume $V=L\times L \times H$. 
We measure the average number of polymer in the bulk, $\langle N_p^\bulk \rangle$, and in the confined system,
$\langle N_p \rangle$. 
We then perform two separate thermodynamic integrations (in the canonical ensemble) to
obtain the free energy of the bulk system with $N_c^\bulk $ colloids
and $\langle N_p^\bulk \rangle$ polymers in a volume $V$, and the confined system of
volume $V$ with $N_c^\bulk$ colloids and $\langle N_p \rangle$ polymers.
In the canonical ensemble simulations, we determined the chemical potential
of the polymer as a consistency check. Typical number of the colloids
and the polymers are $N_c^\bulk=54-900$ and $N_p^\bulk=0-20000$, while
the volume of the simulation box is about $V=(1200-3000) \sigma_c^3$
and $H>16\sigma_c$.  The errors are estimated calculating the standard
deviation from 4 or 5 independent simulations.

\subsection{Adsorption at a hard wall from simulation}
\label{S:simads}
To study the (excess) adsorption of colloid-polymer mixtures at a
planar hard wall we simulated both the bulk mixture and the mixture in
contact with the hard wall in two independent Monte Carlo simulations
in the grand canonical ensemble ensemble and hence we considered only statepoints
of low colloid packing fraction $\eta_c$. 
After discarding 50000 MC steps per particle for equilibration, we take the average of the
number of particles for another 50000 MC steps per particle.  The
differences in particle numbers (per unit area) in the confined system
and in the bulk system then give the adsorption of both species via
Eqs.\ (\ref{Gammac}) and (\ref{Gammap}).

\subsection{Adsorption at a hard wall from scaled-particle theory}
\label{S:spt} 
For a system of hard spheres the scaled particle theory
\cite{HR59,HR60} describes quite accurately the pressure $p$, the hard
wall-fluid interfacial tension $\gamma_\hs$, and the (excess)
adsorption $\Gamma_\hs$, given through the expressions
\begin{eqnarray}
\frac{\beta p}{\rho_c} &=& \frac{1+\eta_c+\eta_c^2}{(1-\eta_c)^3},
\\
\beta \gamma_\hs \sigma_c^2 &=& 3 \eta_c \frac{ (2+\eta_c)}{2 \pi(1-\eta_c)^2},
\label{Eq:spt}
\\
\Gamma_\hs\sigma_c^2 &=& \frac{9 \eta_c^2 }{\pi  (1+ 2 \eta_c)}
-\frac{3 \eta_c}{\pi}.
\label{Eq:gammaHS_spt}
\end{eqnarray}
In particular, Eq.\ (\ref{Eq:gammaHS_spt}) was shown to compare well
with simulation \cite{JRH84} and DFT \cite{RR00} results.

Recently an SPT expression for the wall-fluid tension of AOV model
colloid-polymer mixtures was derived by Wessels \et \cite{PPFW04a}
using the bulk free energy for a ternary mixture obtained from free
volume theory \cite{HNWL+92} as an input, and taking the limit of
vanishing concentration and infinite size of the third
component. Their expression reads
\begin{equation}
\beta \gamma_\wf\sigma_c^2=
\beta \gamma_\hs\sigma_c^2+\eta_p^r f(\eta_c),
\label{E:gspt}
\end{equation}
where $f(\eta_c)= 3 \alpha(\eta_c)/(q^2 \pi) [ 1+(1+3q+q^2)\tau +(3q+4q^2)\tau^2 +3q^2\tau^3]$, 
$\tau=\eta_c/(1-\eta_c)$ and $\beta \gamma_\hs \sigma_c^2$ is given by equation (\ref{Eq:spt}). 
The  polymer free volume is given by the scaled
particle theory  as $\alpha(\eta_c)=  (1 - \eta_c) \exp(-( 3 q+ 3q^2 + q^3) \tau - ( 9q^2/2  +3 q^3 )\tau ^2 - 3 q^3 \tau^3) $. 
Results for $\gamma_\wf$ from equation (\ref{E:gspt}) were found in
Ref.\ \cite{PPFW04a} to compare reasonably well with those from full
numerical density functional calculations. Below we will compare these
approaches against our simulation data.

Here we derive an SPT expression for the adsorption of the AOV model
at a hard wall starting from equation (\ref{E:gspt}) and building
derivatives according to (\ref{E:ad}).  The colloid chemical potential
obtained from the free volume theory \cite{HNWL+92} is
\begin{equation}
\beta \mu_c = \beta \mu_\hs(\eta_c)-\eta_p^r
\frac{\alpha^{'}}{q^3}  , \label{E:mufv}
\end{equation}
where $\beta \mu_\hs(\eta_c)=\eta_c(14 - 13\eta_c + 5\eta_c^2)/(2(1 -
\eta_c)^3) - \log(1 - \eta_c)+ \log( 6\eta_c /\pi ) $ is the SPT
expression of the chemical potential of a system of pure hard spheres
at packing fraction $\eta_c$ and $\alpha^{'}=\partial \alpha /
\partial \eta_c$.  We compute the colloidal adsorption using equation
(\ref{E:ad})
\begin{equation}
 \Gamma_c \sigma_c^2=- \frac{\partial \beta
\gamma_\wf(\mu_c,\mu_p) \sigma_c^2}{\partial \beta \mu_c}=-
\frac{   \partial \beta \gamma_\wf(\eta_c,\mu_p)
\sigma_c^2}{\partial \eta_c} \frac{\partial \eta_c}{\partial \beta
\mu_c},
\end{equation}
where
\begin{equation}
\frac{\partial \eta_c(\mu_c,\mu_p)}{\partial \beta \mu_c}=  \left(
\frac{\partial \beta \mu_c(\eta_c,\mu_p)}{\partial
\eta_c}\right)^{-1},
\end{equation}
is computed using equation (\ref{E:mufv}). The final expression is
\begin{equation}
 \Gamma_c\sigma_c^2=  \Gamma_\hs \sigma_c^2\left( \frac{
1+\eta_p^r \frac{f^{'}}{\beta \gamma_\hs^{'} \sigma_c^2} } {
1-\eta_p^r \frac{\alpha^{''}}{\beta \mu_\hs^{'} q^3}   }  \right) , 
\label{E:gammc}
\end{equation}
where   $f^{'}=\partial f / \partial \eta_c$,
$\gamma_\hs^{'}=\partial \gamma_\hs / \partial \eta_c$,
$\mu_\hs^{'}=\partial \mu_\hs / \partial \eta_c$ and
$\alpha^{''}=\partial^2 \alpha / \partial \eta_c^2$. We note that
the hard-sphere limit is obtained correctly for $\eta_p^r=0$. 
We also calculate the polymer adsorption
\begin{equation}
\Gamma_p \sigma_c^2 =-\frac{\partial \beta
\gamma_\wf(\eta_c,\mu_p)\sigma_c^2}{\partial \beta
\mu_p}=-\eta_p^r f - \left(  \frac{\partial \beta
\gamma_\hs(\eta_c) \sigma_c^2}{\partial \eta_c} + \eta_p^r
\frac{\partial f(\eta_c)}{\partial \eta_c}  \right) \frac{\partial
\eta_c}{\partial \beta \mu_p} .
\end{equation}
Rewriting   Eq.\ (\ref{E:mufv}) as
\begin{equation}
\eta_p^r =\frac{ \beta \mu_\hs(\eta_c)-\beta
\mu_c}{\alpha^{'}}q^3 ,
\end{equation}
we arrive at
\begin{equation}
\frac{\partial \eta_c}{\partial \beta \mu_p}=  \eta_p^r\left(
\frac{\partial \eta_p^r}{\partial \eta_c}\right)^{-1} =
\frac{\eta_p^r \alpha'}{q^3 \beta \mu_\hs' - \eta_p^r
\alpha^{''}}.
\end{equation}
The final expression reads
\begin{equation}
 \Gamma_p(\eta_c) \sigma_c^2=-(\beta \gamma_\wf - \beta \gamma_\hs)\sigma_c^2 +\eta_p^r \frac{ \alpha^{'}}{q^3}  \Gamma_c(\eta_c)\sigma_c^2.
\end{equation}

\subsection{Liquid-gas interfacial tension from capillary wave broadening}
\label{S:capwave} 
Capillary wave theory describes the broadening of an intrinsic
interface of width $w_0$ due to thermal fluctuations.  This broadening
depends primarily on the interfacial tension and the area of the
interface. To calculate the capillary wave contribution to the
interfacial width one has to sum over the contributions from each
thermally excited capillary wave to the amplitude of the oscillations
in the instantaneous interface position.  Here we briefly sketch the
derivation of Sides \et and Lacasse \et and refer the reader to Refs.\
\cite{sides,lacasse} for further details; a very recent study devoted
to capillary waves in colloid-polymer mixtures is that by Vink \et
\cite{RLCV05a}.  Fluctuations due to capillary waves in $\zeta(x,y)$,
the mean location of the interface in the $z$-direction, have an
energy cost due to the increase in surface area of the interface.  The
free energy cost of the interfacial fluctuations is the product of the
excess area of the undulated interface over that of the flat one, and
a liquid-gas interfacial tension $\gamma_{\lg}$, which is assumed to
be independent of curvature. The interfacial Hamiltonian, assuming
that $\zeta$ and its derivatives are small, reads
\begin{equation}
{\cal H} = \frac{\gamma_{\lg}}{2} \int dx dy |\nabla \zeta(x,y)|^2.
\end{equation}
Introducing a Fourier expansion of $\zeta(x,y)$, one arrives at
\begin{eqnarray}
\cal{H}& = &\frac{\gamma_{\lg}}{2} \int d\vec{q} q^2 |
\tilde{\zeta}(\vec{q})|^2,
\end{eqnarray}
where $\vec{q} = (q_x,q_y)$ denotes a two-dimensional wavevector
and $\tilde{\zeta}(\vec{q})$ is the Fourier transform of
$\zeta(x,y)$. Using the equipartition theorem, the mean-square
amplitude for each interfacial excitation mode reads
\begin{equation} \langle |\tilde{\zeta}(\vec{q})|^2\rangle =
\frac{k_BT}{\gamma_{\lg} q^2}.
\end{equation}
The mean-squared real space fluctuations can be calculated by
summing over all allowed modes:
\begin{eqnarray}
\langle |\zeta(x,y)|^2\rangle &=& \frac{1}{(2 \pi)^2}
\int_{q_{\rm min}}^{q_{\rm max}} d \vec{q} \langle
|\zeta(\vec{q})|^2\rangle \nonumber \\
& = & \frac{k_BT}{2 \pi \gamma_{\lg}} \ln \left ( \frac{L}{\xi_b}
\right ),
\end{eqnarray}
where the  low $q$ cutoff, $q_{\rm min}$, is determined by the system
size, i.e., $q_{\rm min}=2 \pi/L$ in our simulations and gives rise to
system size dependence. The high $q$ cutoff, $q_{\rm max} = 2
\pi/\xi_b$, is determined by the bulk correlation length $\xi_b$,
which is of the order of the colloid diameter, and avoids the
divergence in the integral. 

The total width of the interface, as measured in experiments and
simulations, includes contributions from the intrinsic width and
the broadening due to capillary wave fluctuations. If one assumes
that the capillary-wave fluctuations are decoupled from the
intrinsic profile, the total interfacial profile $\Psi(z)$ can be
expressed as a convolution of the intrinsic interfacial profile
$\psi(z)$ and the fluctuations due to capillary waves
\begin{equation}
\Psi(z) = \int_{-\infty}^{\infty} \psi(z-z_0) P(z_0) dz_0,
\end{equation}
where $P(z_0)$ is the probability of finding the local interface
at $z_0$. The interfacial order parameter profile $\Psi(z)$ is
defined such that it varies between -1 and 1
\begin{equation}
\Psi(z) = \frac{2}{\rho_c^l-\rho_c^g}\left\lbrack \rho_c(z) -
\frac{\rho_c^l+\rho_c^g}{2}\right \rbrack,
\label{Eq:intpar}
\end{equation}
where $\rho_c(z)$ is the cross-section averaged density profile of
the colloids and $\rho_c^l$ and $\rho_c^g$ are the colloid
densities of the colloidal "liquid" and "gas" phase at
coexistence. In Ref.\ \cite{sides,lacasse}, the authors define the
variance of the derivative of the total interfacial profile
$d\Psi(z)/dz \equiv \Psi'$ as a measure of the width of the
interface. The variance of a distribution $f$ reads
\begin{equation}
v\lbrack f\rbrack = \frac{\int_{-\infty}^{\infty} z^2 f(z)
dz}{\int_{-\infty}^{\infty} f(z) dz} =
\frac{-(d^2/dq^2)\tilde{f}(q)|_{q=0}}{\tilde{f}(0)},
\end{equation}
where $\tilde{f}(q)$ is the Fourier transform of $f(z)$. Using
this choice of measure for the interfacial width, one can show
explicitly using the convolution theorem that the total
interfacial width can be written as the sum of an intrinsic part
and a contribution due to capillary wave fluctuations
\begin{eqnarray}
v\lbrack \Psi' \rbrack & = & v \lbrack \psi' \rbrack + v \lbrack P
\rbrack \nonumber \\
& = & v \lbrack \psi' \rbrack + \frac{k_BT}{2 \pi \gamma_{\lg}} \ln
\left (\frac{L}{\xi_b} \right ),
\label{sizedep}
\end{eqnarray}
where we identify $v \lbrack P \rbrack$ as the mean-squared
fluctuations due to capillary waves, i.e., $\langle
|\zeta(x,y)|^2\rangle$.

\begin{figure}[p]
\begin{center}
\includegraphics[width=8cm]{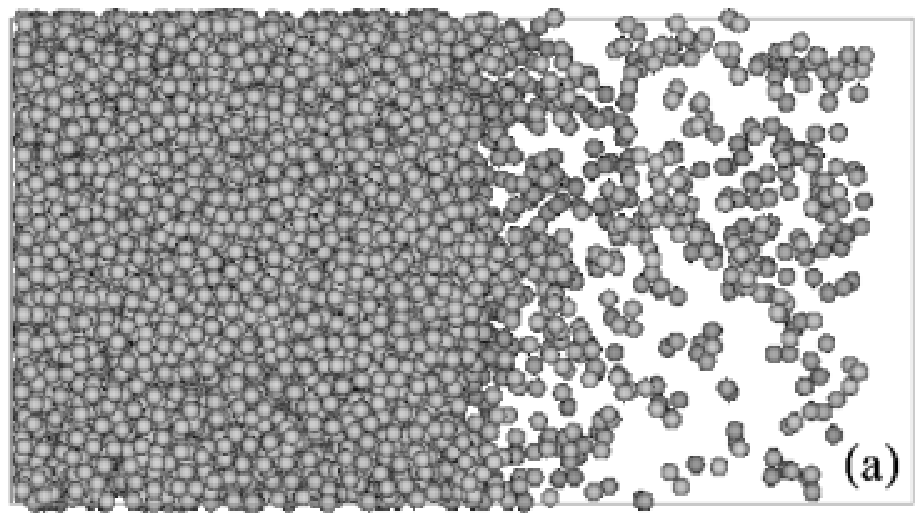}
\includegraphics[width=8cm]{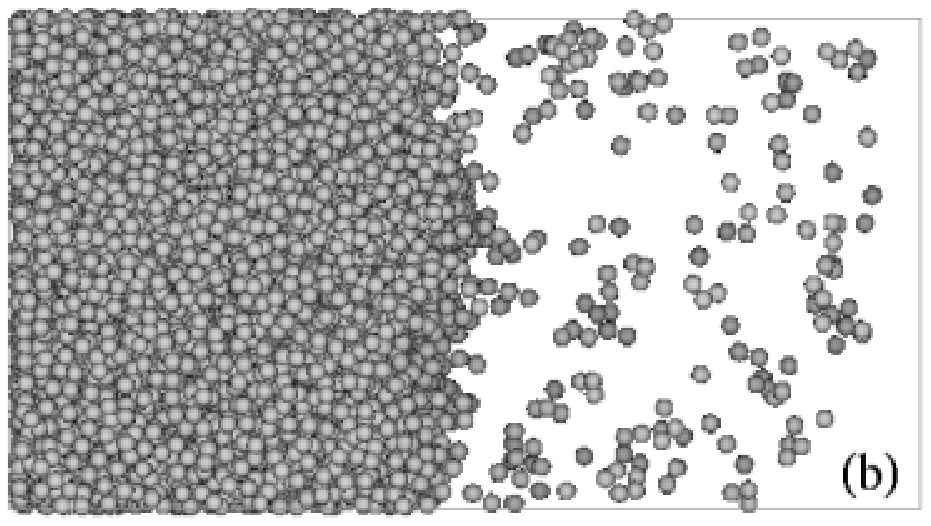}
\includegraphics[width=8cm]{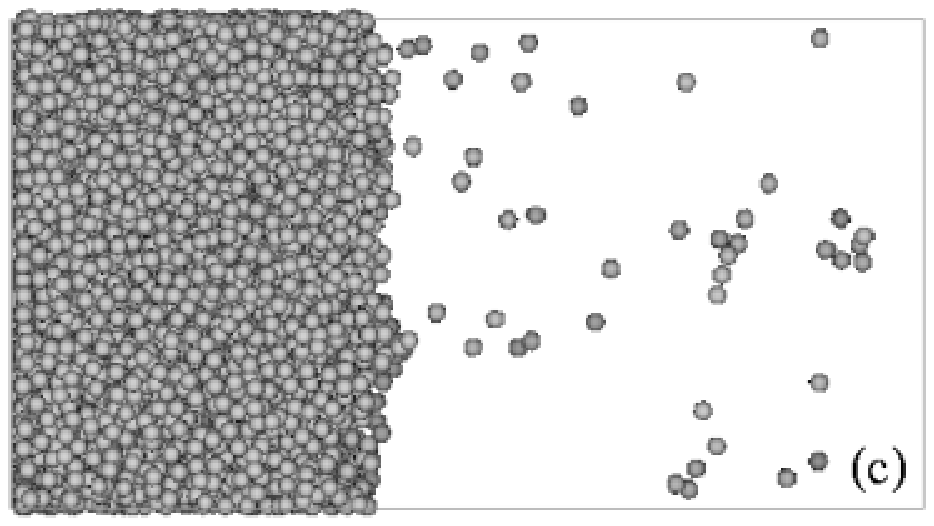}
\includegraphics[width=8cm]{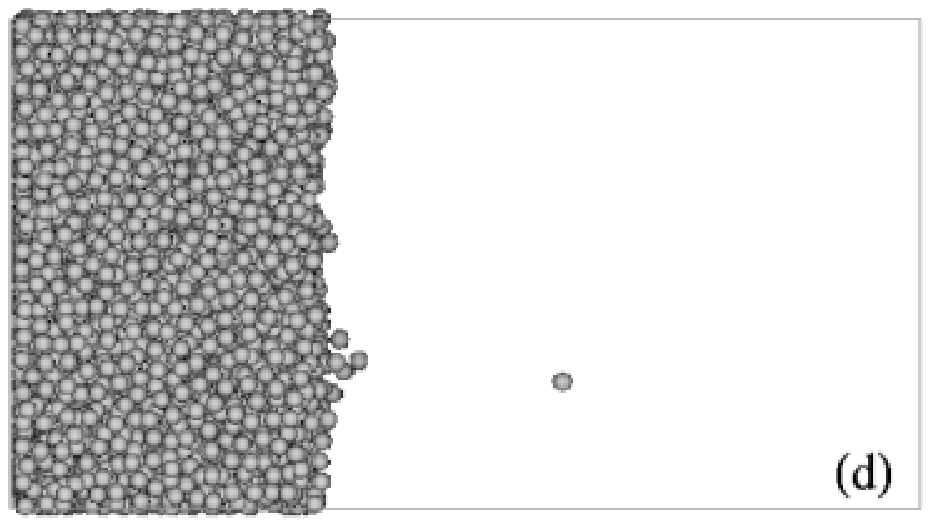}
\end{center}
\caption{Typical snapshots of the colloid coordinates, obtained from Monte-Carlo simulations based on the exact effective one-component Hamiltonian for the colloids, of the liquid-gas interface of the model
colloid-polymer mixture with size ratio $q=1$ and a) polymer reservoir
packing fraction $\eta_p^r=0.95$; b) $\eta_p^r=1.05$; c)
$\eta_p^r=1.4$; d) $\eta_p^r=2$. }
\label{snap}
\end{figure}

To measure the tension using the results of capillary wave
theory one needs to create a liquid-gas interface in the simulation
box.  To stabilize the liquid-gas interface we perform the simulations
in a box with dimensions $L \times L \times H$, with $H = 48
\sigma_c$, delimited in the $z$ direction by one impenetrable wall and
one semi-permeable wall. We vary the lateral dimensions in the range
of $5 \sigma_c<L<25 \sigma_c$. The canonical simulations are started
in the middle of the two-phase region. After equilibration the system
is phase separated, the gas phase in contact with the semi-permeable
wall and the liquid phase in contact with the hard-wall. The
liquid-gas interface is in the middle of the simulation box (see also
Fig. \ref{snap}).  We determine the total interfacial width $v\lbrack
\Psi' \rbrack$ from the interfacial order parameter profile $\Psi(z)$
measured in simulations according to Eq. (\ref{Eq:intpar}).
A priori, it is not clear whether the density profile should be fit to
an error function or a hyperbolic tangent.  In Ref.\ \cite{sides}, it
is shown explicitly that the interfacial width extracted through fits
of the density profile to a hyperbolic tangent lead to systematic
errors and that better results for the interfacial tensions are
obtained using the error function. In addition, later work on
water/carbon tetrachloride \cite{senapati} and molten salt (KI)
\cite{scott} interfaces also found good agreement between the
interfacial tension calculated using the capillary wave formalism and
the pressure tensor components.
Fitting $\Psi(z)$ by an error function $\mbox{erf}\lbrack(z-z_0)/(w
\sqrt{2})\rbrack$, using $z_0$ and $w$ as fitting parameters, we
find that the variance of the derivative of this fitting function
is related to the interfacial width $w$, i.e.,  $v\lbrack \Psi'
\rbrack = w^2$. Using Eq.\ (\ref{sizedep}) we are able to determine
$\gamma_{\lg}$ from the fits of the size dependence of the
interfacial width.

In contrast to the wall-fluid tension  and the adsorption
simulations described in section \ref{S:simwf} and \ref{S:simads}, 
we used for the liquid-gas interfacial tension simulations described in this section and in section \ \ref{S:probabilityDistribution}, an efficient simulation scheme for the AOV model that
was recently developed \ \cite{MDRvR02}. It is based on the exact effective
one-component Hamiltonian of the colloids, i.e., it incorporates
all the effective polymer-mediated many-body interactions. The
effective one-component Hamiltonian can be derived by integrating
out the polymer degrees of freedom in the binary colloid-polymer
mixture. To this end, we consider $N_c$ colloids and $N_p$ polymer
coils in a macroscopic volume $V$ at temperature $T$. The total
Hamiltonian consists of interaction terms
$H=H_{cc}+H_{cp}+H_{pp}$, where $H_{cc}=\sum_{i<j}^{N_c}
v_{cc}(R_{ij})$, $H_{cp}=\sum_{i}^{N_c}\sum_{j}^{N_p} v_{cp}(|{\bf
R}_{i}-{\bf r}_j|)$, and $H_{pp}=\sum_{i<j}^{N_p} v_{pp}(r_{ij})$.
It is convenient to consider the system in the $(N_c,z_p,V,T)$
ensemble, in which the polymer fugacity is fixed. The
thermodynamic potential $F$ of this system can be written as
$\exp\lbrack - \beta F \rbrack = \mbox{Tr}_c \exp \lbrack - \beta H_{eff}
\rbrack$, where $H_{eff} = H_{cc} + \Omega$ is the effective
Hamiltonian and where $\mbox{Tr}_c$ is short for $1/N_c!\Lambda_c^{3N_c}$
times the volume integral over the coordinates of the colloids,
and where $\Lambda_c$ is the thermal wavelength. It is
straightforward to show that one obtains for the present model the
exact result $\beta \Omega = -z_pV_f$, with the so-called free
volume $V_f$, i.e., the accessible volume for the center-of-mass
of the polymer coils. This free volume can be calculated
numerically on a smart grid for each static colloid configuration.
For more details, we like to refer the reader to Ref. \ \cite{MDRvR02,MD05}.
The advantage of this scheme is that the polymer degrees of
freedom are integrated out and enter the effective one-component
colloid Hamiltonian only by the polymer reservoir packing fraction
$\eta_p^r$. Hence, we avoid equilibration and statistical accuracy
problems due to fluctuating polymer numbers. Moreover, our
simulations are not limited by the total number of polymers and
can be performed at high polymer reservoir packing fractions far
away from the critical point.

\subsection{Liquid-gas interfacial tension from the probability distribution}
\label{S:probabilityDistribution}
In addition, we determine $\gamma_{\lg}$ using the probability
distribution method. We perform Monte-Carlo simulations in the
grand canonical ensemble using again our novel effective
one-component simulation scheme \cite{MDRvR02}, explained in Sec. \ \ref{S:capwave} . 
To obtain the probability $P(N_c)|_{z_c,\eta_p^r}$ of observing $N_c$
colloids in a volume $V$ at fixed colloid fugacity $z_c$ and fixed
polymer reservoir packing fraction $\eta_p^r$, we use a sampling
technique called successive umbrella sampling \cite{virnau}.
Employing this technique, we sample the probability distribution
$P(N_c)|_{z_c,\eta_p^r}$ in small windows one after the other, in
which the number of colloids $N_c$ is allowed to fluctuate between
0 and 1 in the first window, 1 and 2 in the second window, etc. We
first perform an exploratory short run without a bias, yielding
$P(N_c-1)$ and $P(N_c)$ for window $N_c$. We then perform a biased
simulation in which we sample from a non-physical distribution
$\pi(N_c)=g(N_c)P(N_c)$ instead of the grand canonical
distribution $P(N_c)$. We have chosen the weight function
$g(N_c)=1/P(N_c)$, where we use $P(N_c)$ obtained from the
unbiased exploratory simulation run. This choice for the weight
function yields a constant constant $\pi(N_c)$ and the system will visit
equally  the state with $N_c-1$ colloids as the state with $N_c$
colloids. Of course, we have to correct our weighted sampling for
the bias by dividing out the weight function $g(N_c)$. The more
accurate grand canonical distribution $P(N_c)$ is obtained from
\begin{equation}
P(N_c) = \frac{\pi(N_c)}{g(N_c)}.
\end{equation}
In addition, we use the histogram reweighting technique to obtain
the probability distribution for any $z_c'$ once
$P(N_c)|_{z_c,\eta_p^r}$ is known for a given $z_c$ \cite{RLCV05}:
\begin{equation}
P(N_c)|_{z_c',\eta_p^r} = \ln P(N_c)|_{z_c,\eta_p^r} +
\ln\left(\frac{z_c'}{z_c}\right)N_c \label{histogram}.
\end{equation}
At phase coexistence, the distribution function $P(N_c)$ becomes
bimodal, with two separate peaks of equal area for the "colloidal
" liquid and gas phase. To determine phase coexistence,
we normalize $P(N_c)|_{z_c,\eta_p^r}$ to unity
\begin{equation}
\int_{0}^{\infty}P(N_c)|_{z_c,\eta_p^r}dN_c = 1 \label{norm}
\end{equation}
and we determine the average number of colloids 
\begin{equation}
\langle N_c \rangle=
\int_{0}^{\infty}N_cP(N_c)|_{z_c,\eta_p^r}dN_c.
\end{equation}
Using the histogram reweighting technique (\ref{histogram}) we
determine for which $z_c'$ the equal area rule\begin{equation}
\int_{0}^{\langle N_c \rangle }P(N_c)|_{z_c',\eta_p^r}dN_c =
\int_{\langle N_c \rangle }^{\infty} P(N_c)|_{z_c',\eta_p^r}dN_c,
\end{equation}
 representing the condition for phase coexistence, is satisfied.

The liquid-gas interfacial tension $\gamma_{\lg}$ for a finite system of volume
$V = L^2 H$ can be obtained from $P(N_c)|_{z_c',\eta_p^r}$ at
coexistence:
\begin{equation}
\gamma_{\lg,L} =\frac{1}{2 L^2}\left\lbrack
\ln \left(\frac{P(N_{c,\text{max}}^g )+P(N_{c,\text{max}}^l)}{2}\right) -
\ln(P(N_{c,\text{min}}))\right \rbrack
\end{equation}
where $P(N_{c,\text{max}}^g)$ and $P(N_{c,\text{max}}^l)$ are the
maxima of the gas and liquid peaks, respectively, and
$P(N_{c,\text{min}})$ is the minimum between the two peaks. We can 
determine the interfacial tension for the infinite system, i.e.,
$\gamma_{\lg}$, by performing simulations for a range of systems
sizes and by extrapolating the results to the infinite system, as shown by Binder, using the relation\ \cite{KB82,JJP00,RLCV05}
\begin{equation} \gamma_{\lg,L} =
\gamma_{\lg} - \frac{x\ln L}{2 L^2}- \frac{\ln A}{2 L^2},
\label{Eq:fit1}
\end{equation}
where $A$ and $x$ are generally unknown. 

\subsection{Liquid-gas interfacial tension from Young's equation}
\label{S:young}
Young's equation relates $\gamma_{\lg}$ to the difference in wall-fluid
tension for the gas, $\gamma_{\rm wg}$, and the liquid phase,
$\gamma_{\rm wl}$, via
\begin{equation}
 ( \gamma_{\rm wg} - \gamma_{\rm wl})=\gamma_{\lg} \cos\theta,
\end{equation}
where $\theta$ is the (macroscopic) contact angle at which the
gas-liquid interface hits the wall.  In the region of complete
wetting, the contact angle is zero and hence $\gamma_{\lg}$ can be
obtained from the difference of the wall-gas and wall-liquid tensions,
$\gamma_{\lg}= (\gamma_{\rm wg} - \gamma_{\rm wl})$.

\subsection{Density functional theory for interfacial properties}
\label{S:dft} 
We use the approximation for the Helmholtz excess free energy for the
AOV model as given in \cite{MSHL+00}. For given external potential,
the density functional is numerically minimized using a standard
iteration procedure. The interfacial tension is then obtained from
Eq.\ (\ref{fgamma}), and the adsorption of both species from Eqs.\
(\ref{Gammac}), (\ref{Gammap}). Technical details about the DFT
implementation that also apply to the present study are given in
\cite{PPFW04}.

\section{Results}
\label{S:res}

\subsection{Wall-fluid interfacial tension}
\label{S:reswf}
\begin{figure}
  \begin{center}
\includegraphics[width=14cm]{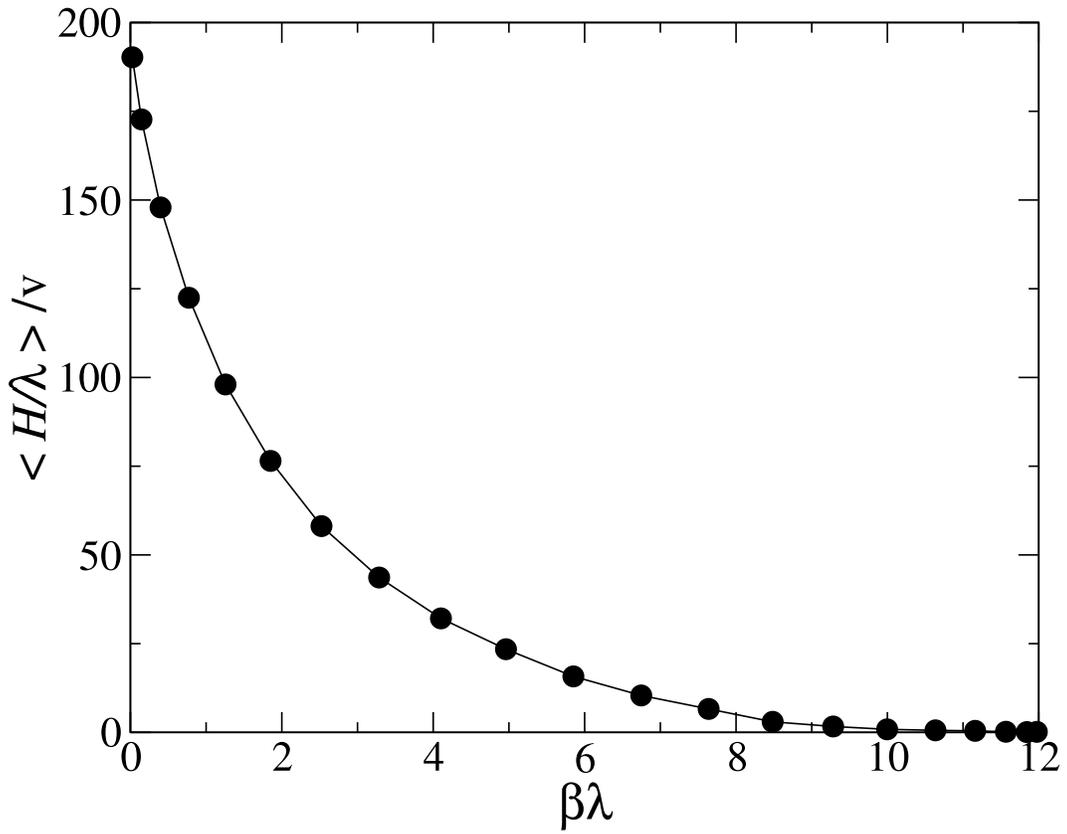}
\caption{The average number of overlaps $ \left \langle \frac{H}{ \lambda} \right \rangle_{\lambda} $ ,
 between colloidal particles and the walls per unit volume as a
 function of the height of the step potential for the pure colloidal
 system ($\eta_p^r=0$) with packing fraction $\eta_c=0.4$. For $\beta
 \lambda \geq 10$ the number of overlaps is zero within statistical
 fluctuations and the system behaves like the hard-sphere system. The
 line is a guide to the eye.}
    \label{FIG:integrand}
  \end{center}
\end{figure}

\begin{figure}
  \begin{center}
\includegraphics[width=14cm]{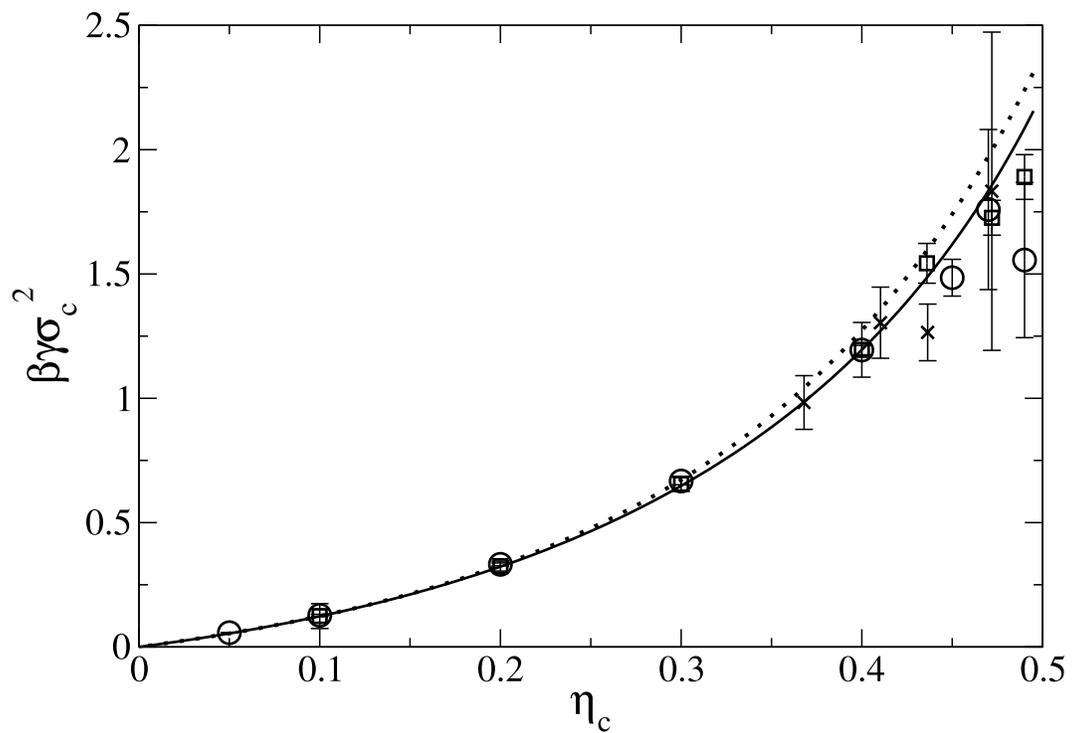}
\caption{The reduced wall-fluid interfacial tension $\beta \gamma
\sigma^2$ of hard spheres adsorbed at a hard wall as a function of the
colloidal packing fraction $\eta_c$. We compare our simulation results
(open circles) with Monte Carlo simulations\ \cite{MHHL99} (open squares) and  Molecular Dynamics simulations\ \cite{JRH84} (crosses). The dotted line indicates
the result from SPT and the solid line denotes the DFT result.}
\label{FIG:hs_gamma}
\end{center}
\end{figure}

In this section we present the results on the interfacial tension of model colloid-polymer mixtures from simulation, SPT and DFT. 
We checked the simulation technique performing simulations for a system of pure hard spheres
($\eta_p^r=0$) and started by finding the optimal value for  
$\lambda_{\rm max}$, the maximum height of the potential (\ref{E:ham}). 
To this end, we computed the average number of
overlaps $ \left \langle \frac{ H}{ \lambda} \right \rangle_{\lambda} $ among particles and walls for different values of the
potential height $\lambda$, with $H$ defined by the equation (\ref{E:ham}).
The value of $\lambda_{\rm max}$ depends
on the particle packing fraction. 
In Fig.\ \ref{FIG:integrand} we plot the average number of overlaps between the colloidal
particles and the walls for packing fraction $\eta_c=0.4$. For all
values  of $\beta\lambda \geq 10$, with $\beta=1/kT$ the inverse
temperature, the average number of overlaps is zero within the
statistical fluctuations. Smaller packing fractions require
smaller values of $\beta\lambda_{\rm max}$. 
We checked the reliability of the approximation by computing the reduced wall-fluid interfacial tension
$\beta \gamma_\hs \sigma_c^2$ (Fig.\ \ref{FIG:hs_gamma}) and found that our
simulation results are consistent with the results of Ref.\ \cite{MHHL99}. 
At packing fraction $\eta_c \gtrsim 0.45 $ the
precision of the simulation is low  but comparable with the
simulation method of Heni and L\"owen \cite{MHHL99} using the wall
insertion method, and of Henderson and van Swol using the pressure
tensor method \cite{JRH84}. The agreement between simulation and
density functional theory (DFT) \cite{RR00,PPFW04a} is remarkably good.
The scaled particle theory (SPT) \cite{PPFW04a} overestimates the
wall-fluid tension at high density. This is due to the the
inaccuracy of the SPT equation of state. Better results can be
obtained combining the SPT equation for the interfacial tension
and the Carnahan-Starling equation of state as explained in detail in Ref.\ \cite{MHHL99}.

\begin{figure}[p]
\begin{center}
\includegraphics[width=14cm]{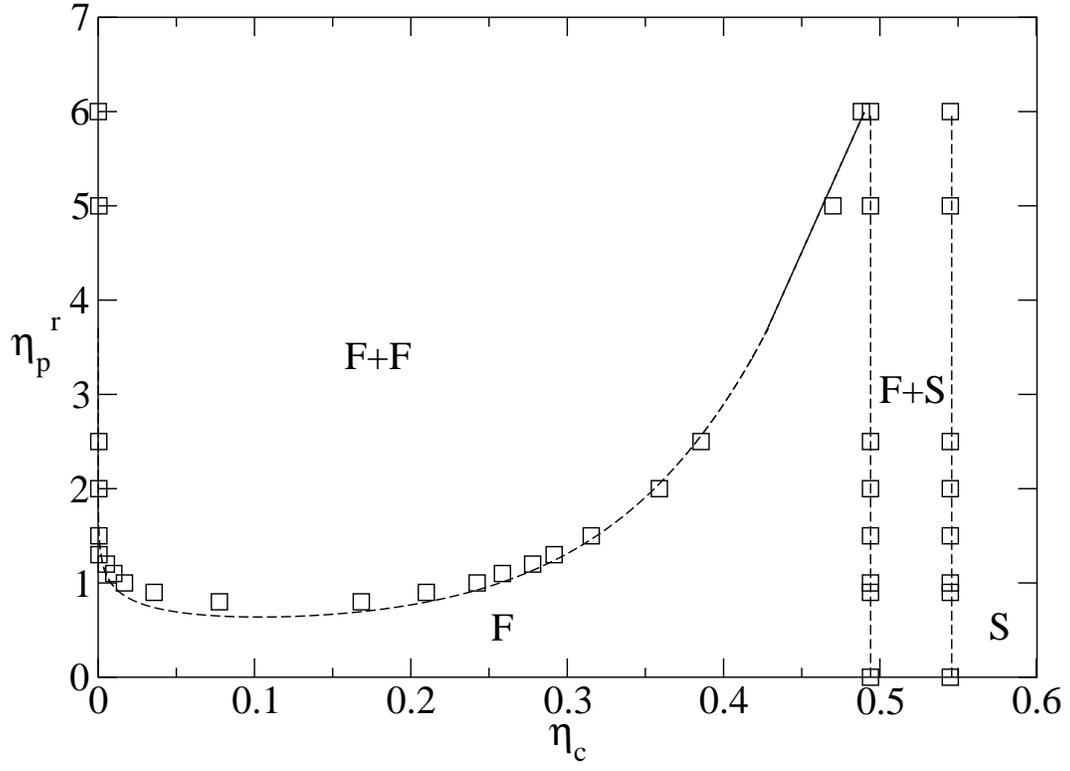}
\end{center}
\caption{Phase diagram of the AOV model for size ratio $q=1$ as
obtained from simulations, taken from Refs.\ \cite{MDRvR02} (symbols), and free volume theory\ \cite{HNWL+92} (dashed line) as a function of the colloid packing fraction $\eta_c$ and the polymer reservoir
packing fraction $\eta_p^r$. F and S denote the stable fluid and solid (fcc) phase. F+S and F+F denote, respectively, the stable fluid-fluid, and fluid-solid coexistence region.}
\label{F:binodal}
\end{figure}

\begin{figure}[p]
\begin{center}
\includegraphics[width=8cm]{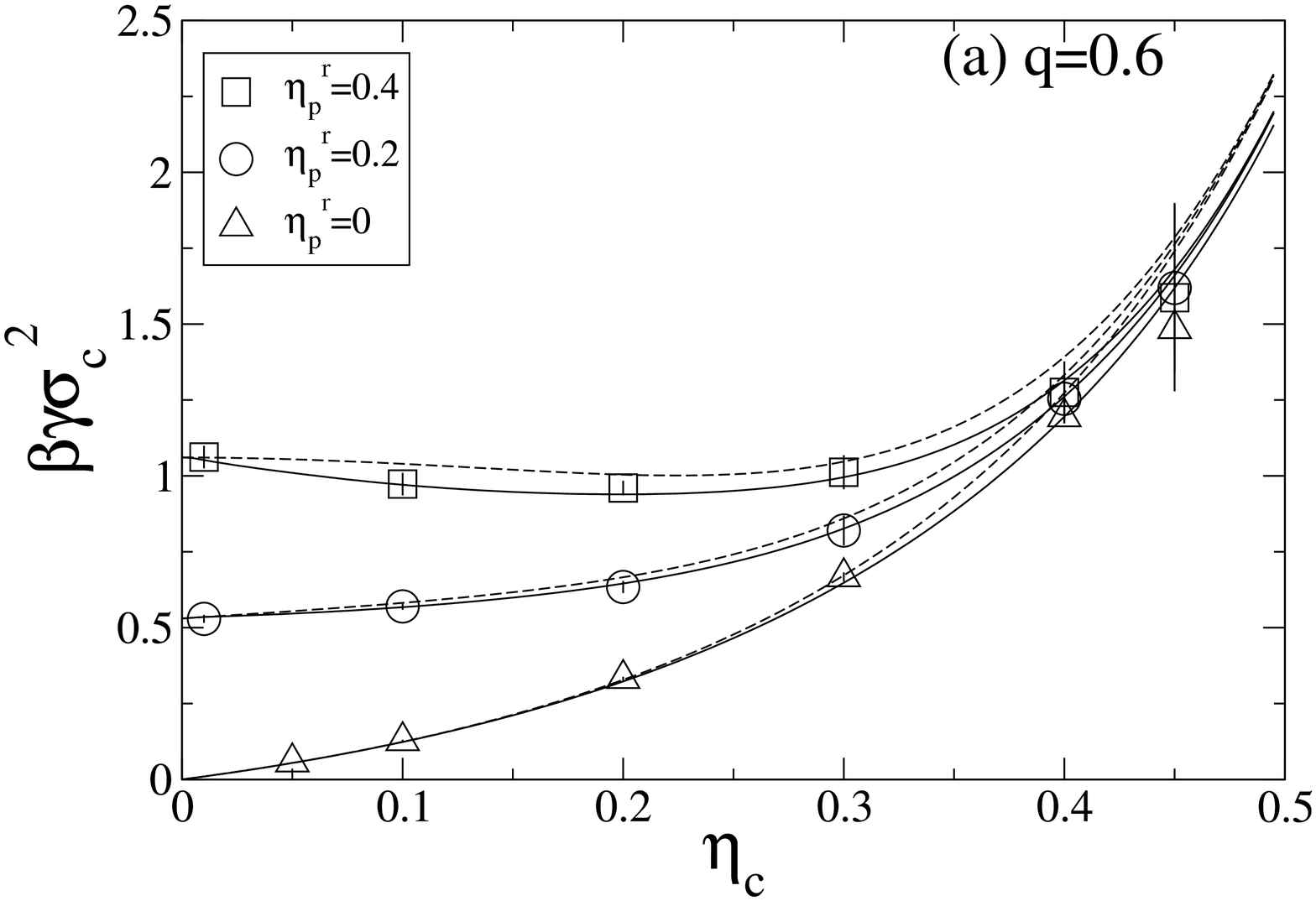}
\includegraphics[width=8cm]{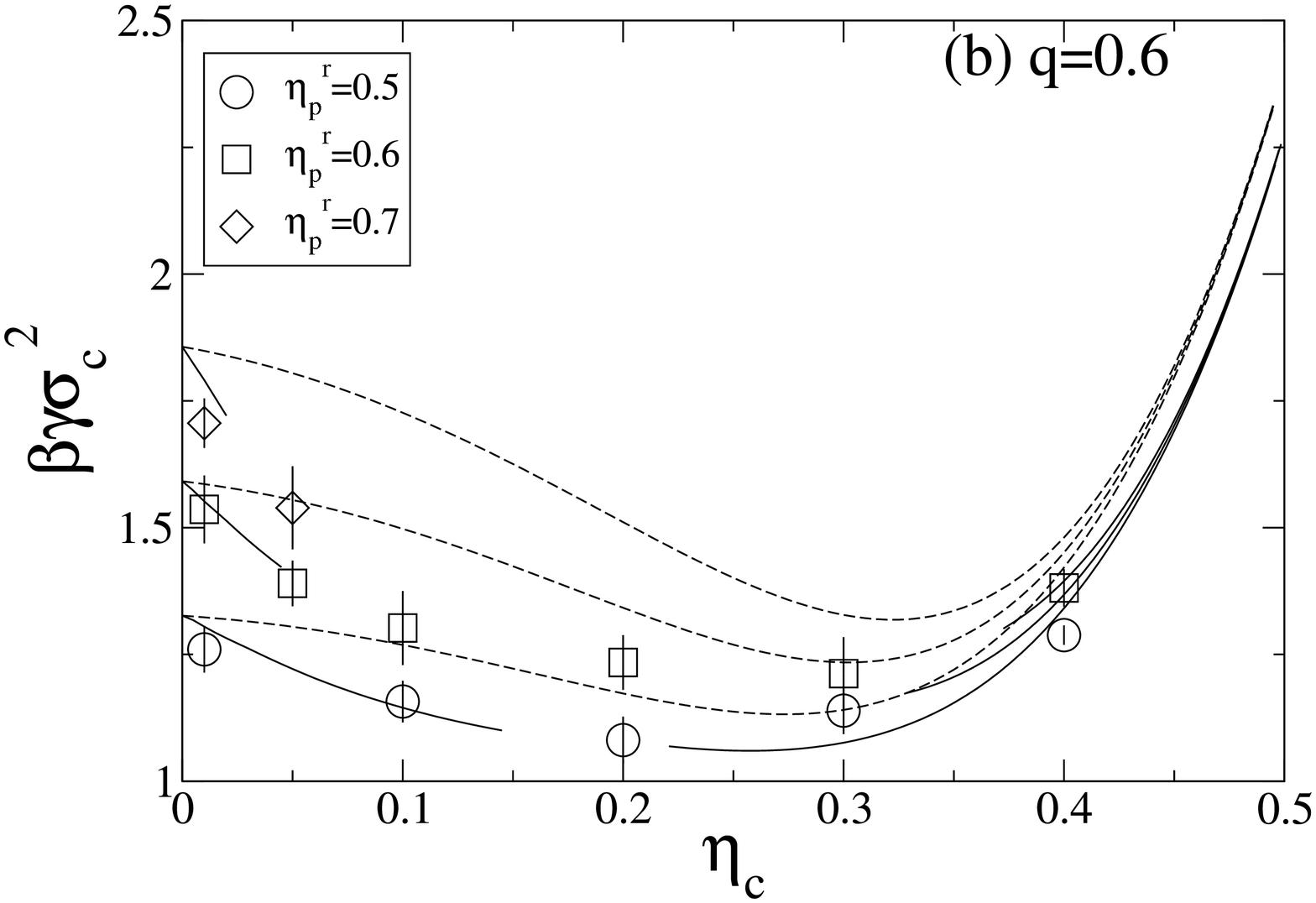}
\includegraphics[width=8cm]{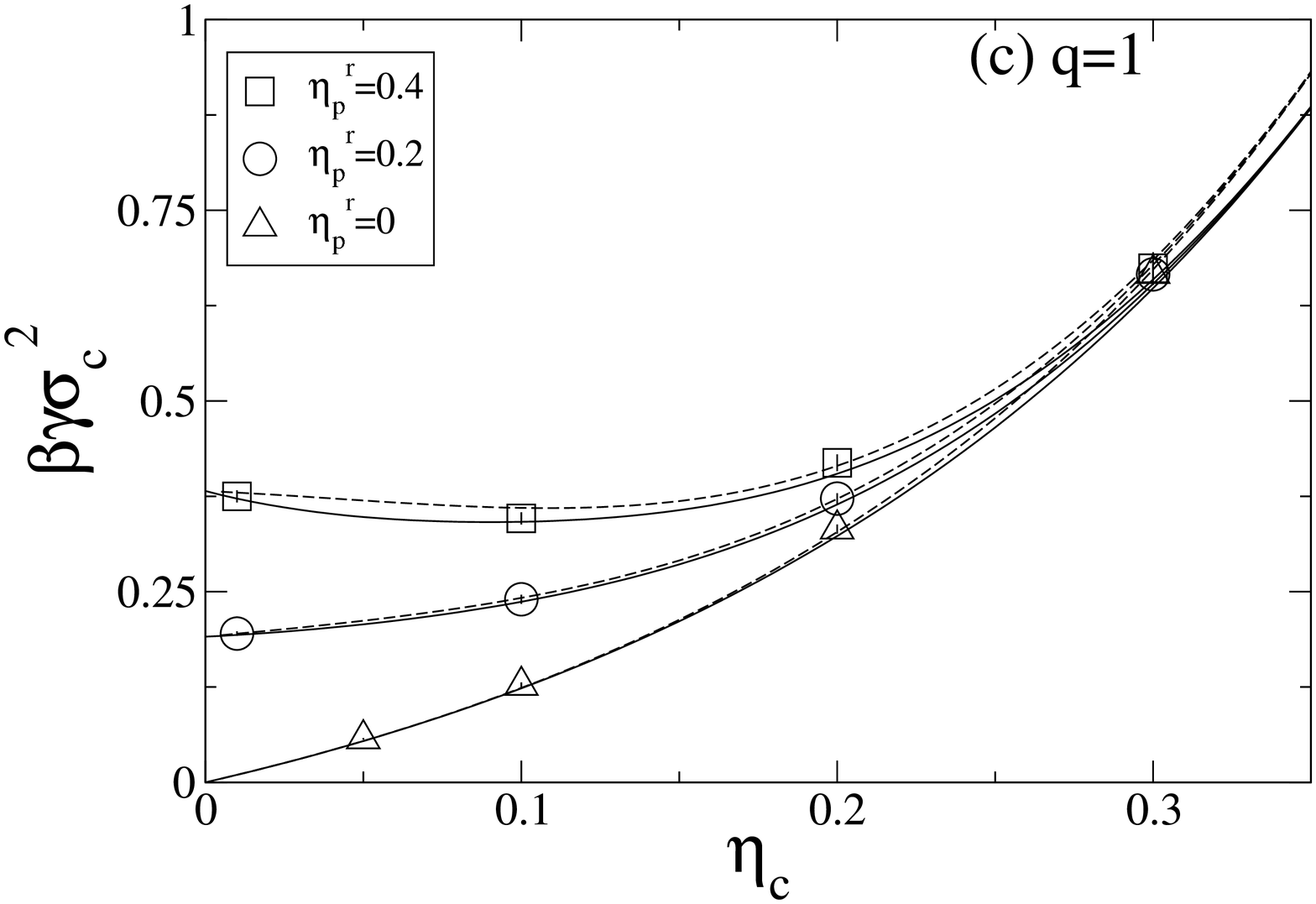}
\includegraphics[width=8cm]{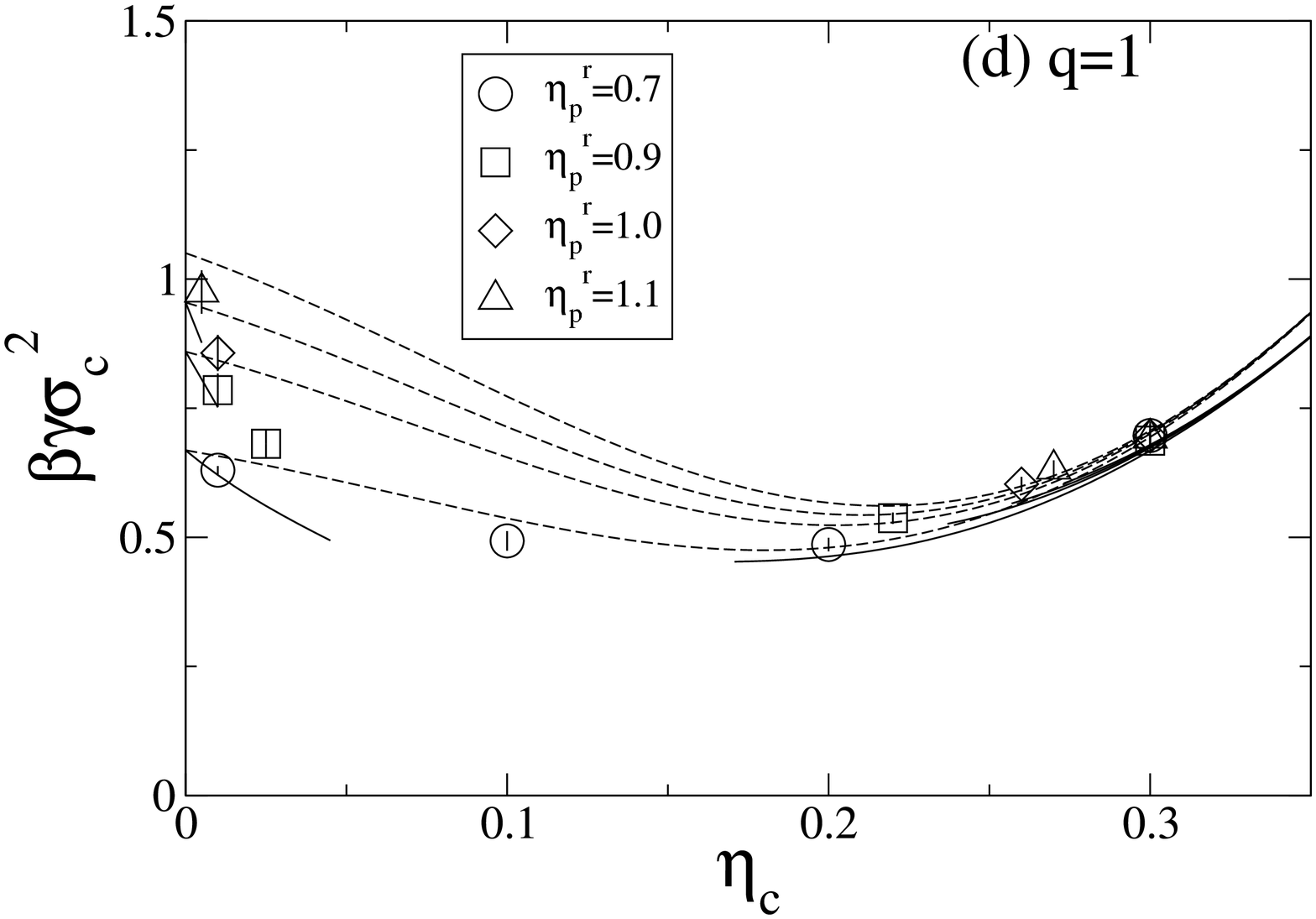}
\end{center}
\caption{The wall-fluid interfacial tension of the model
colloid-polymer mixture adsorbed against a hard wall. The symbols
denote simulation results, dotted curves denote SPT results
\cite{PPFW04a}, the solid curves denote DFT results \cite{PPFW04a}.
a) Size ratio $q=0.6$ and $\eta_p^r=$0, 0.2, and 0.4; 
b) size ratio $q=0.6$ and $\eta_p^r=$0.5, 0.6, and 0.7;
c) size ratio $q=1$ and $\eta_p^r=$0, 0.2, and 0.4;
d) size ratio $q=1$ and $\eta_p^r=$0.7, 0.9, and 1.0.}
 \label{F:gamma}
\end{figure}

We now determine the wall-fluid interfacial tension for AOV colloid-polymer mixtures of size ratio $q=0.6$ and $q=1$
for different values of the polymer reservoir packing fraction $\eta_p^r$ and of the colloid packing fraction $\eta_c$. 
The addition of nonadsorbing polymers to a colloidal suspension of hard-sphere can induce a phase separation.  
In Fig.\ \ref{F:binodal} we show the bulk phase diagram for size ratio $q=1$ from
previous simulations\ \cite{MDRvR02}  in the ($\eta_p^r$,$\eta_c$) representation. For
comparison, we also plot the phase diagram obtained from free volume
theory, which is equivalent to our DFT phase diagram\ \cite{HNWL+92}. At $\eta_p^r=0$
we find the freezing transition of the pure hard-sphere system with
packing fractions $\eta_c^f \simeq 0.494$ and $\eta_c^s \simeq 0.545$ for the
coexisting fluid and solid phase, respectively. The critical point is
estimated to be at $\eta_{p,crit}^r=0.86$, while DFT, equivalent to the
free-volume theory predicts  $\eta_{p,crit}^r$= 0.638. 
For $\eta_p^r < eta_{p,crit}^r$, there is a stable fluid phase for
$\eta_c<0.494$, a fluid-solid coexistence region for $0.494<\eta_c<0.545$, and
a stable solid phase (fcc crystal) for $\eta_c>0.545$. 
For $\eta_p^r>\eta_{p,crit}^r$, a fluid-fluid coexistence region appears where the system demixes in a colloidal liquid phase, rich in colloids and poor in polymers, and a colloidal gas phase, that is poor in colloids and rich in polymers.
The triple point, where the gas, the liquid, and the solid are in
coexistence, is located at $\eta_{p,\rm triple}^r=6$.
For $\eta_p^r>\eta_{p,triple}^r$, the fluid-fluid coexistence region disappears, and a wide crystal-fluid coexistence region appears. 
The overall phase diagram is analogous to that of a simple fluids upon identifying $\eta_p^r$ with the inverse temperature.
Despite differences near the critical
point, DFT and simulations results are in good
agreement for state points at $\eta_p^r>1.5$.   
In Fig.\ \ref{F:gamma}a and Fig.\ \ref{F:gamma}c we show the wall-fluid tension for state points below the
gas-liquid critical point for size ratio $q=0.6$ and $q=1$, respectively. 
For comparison, we also plot the results for pure
hard spheres ($\eta_p^r=0$). 
The addition of non-adsorbing polymers to a suspension of hard-sphere colloids (i.e. increasing
$\eta_p^r$) increases the wall-fluid interfacial tension.
For $\eta_c=0$, the wall-tension is the work done to introduce an
impenetrable wall in an ideal gas of polymers divided by the total
area: $\beta \gamma(\eta_c=0)=\beta P^{id} \sigma_p/2$, where
$\beta P^{id}=\rho_p^r$ is the bulk pressure of the ideal gas
of polymer and $\sigma_p/2$ is the thickness of the depletion
layer of the polymer at the wall. 
For small $\eta_c$, the  slope of the tension is smaller than in the hard sphere case and for $\eta_p^r >=0.4$ it is negative. This is due to the attractive interaction that arises between the colloidal particles and the walls.  
For large $\eta_c$ the interfacial tension approaches that of pure hard spheres  as at
high colloid density the number of polymers in the mixture
rapidly approaches zero. 
Simulations and DFT are in good agreement for all state points that we considered. 
The SPT predicts correctly the value at $\eta_c=0$, but it systematically overestimates the
wall-fluid tensions for all values of $\eta_c>0$. 
One can show that the low $\eta_c$ expansion violates an exact wall sum rule\ \cite{RR1}. 
The deviation increases with increasing $\eta_p^r$. In Fig.\ \ref{F:gamma}b we
show the results for size ratio $q=0.6$ for state points that are
at higher $\eta_p^r$ than the DFT gas-liquid critical point. We
did not calculate the binodal with computer simulation, but the
system was still in the one phase region of the phase diagram for
$\eta_p^r=0.5$ and 0.6. For comparison, we also plot the results
from DFT and SPT. Note that DFT results are only shown in the
stable gas and liquid regimes and are hence disconnected from each
other, showing the biphasic region at intermediate $\eta_c$. In
Fig.\ \ref{F:gamma}d we show the results for the size ratio $q=1$
for state points that are at higher $\eta_p^r$ than  the DFT
gas-liquid critical point. For small $\eta_c$ the SPT fails to
reproduce the slope of the curves, due to the absence of colloid
correlations (layering) near the hard-wall in SPT theory.

\subsection{Adsorption at a hard wall}
\label{S:resads}
\begin{figure}[p]
\begin{center}
\includegraphics[width=8cm]{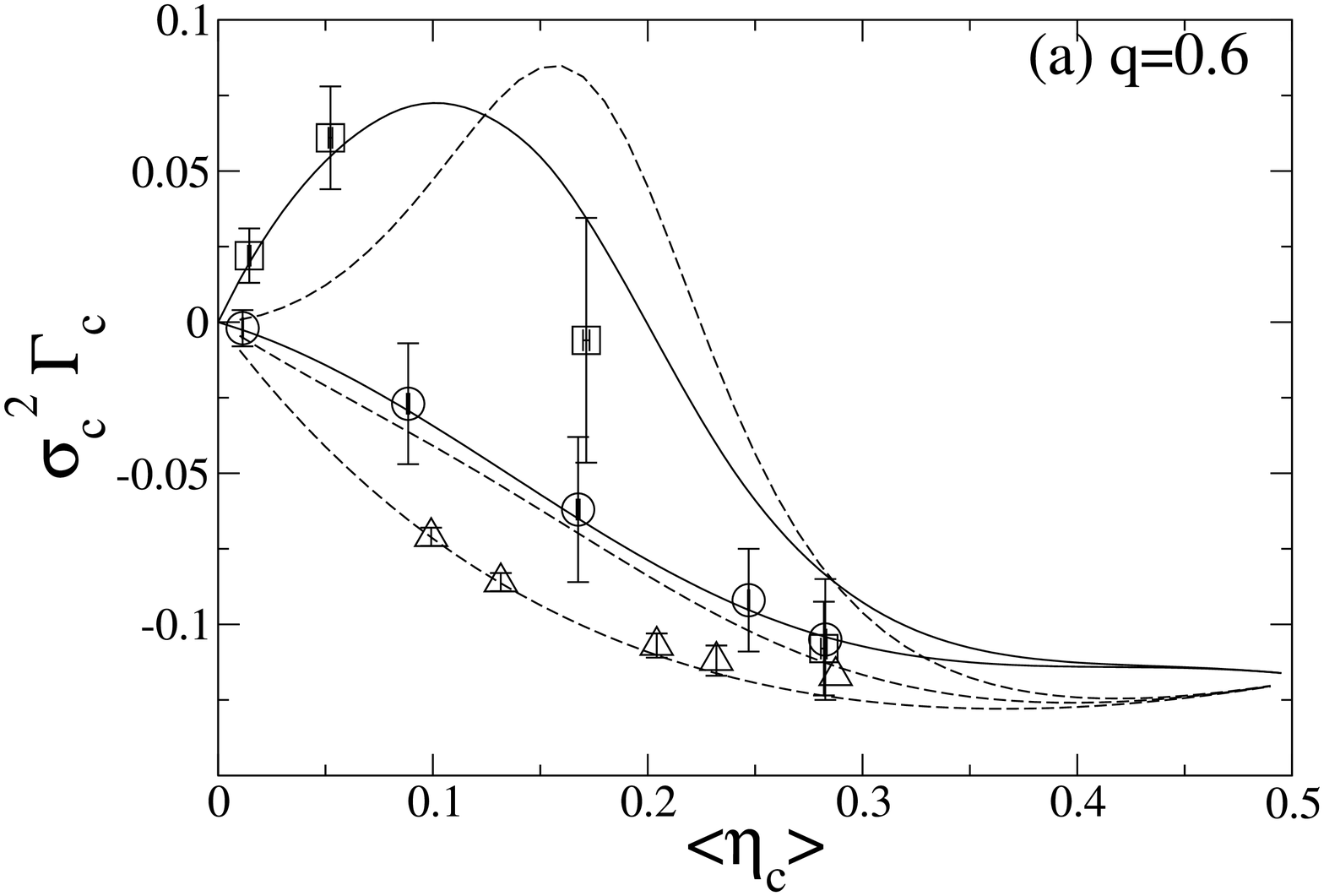}
\includegraphics[width=8cm]{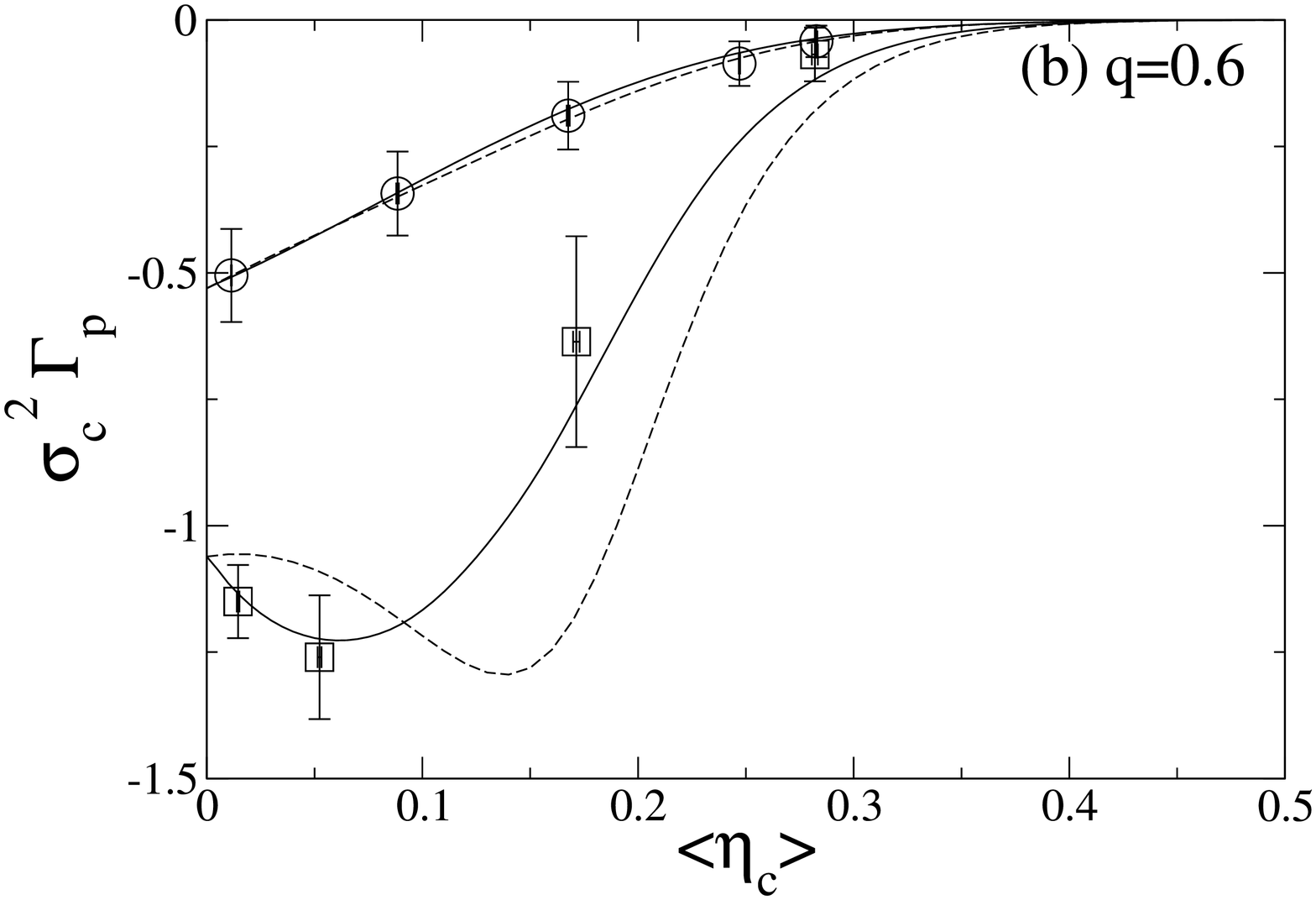}
\includegraphics[width=8cm]{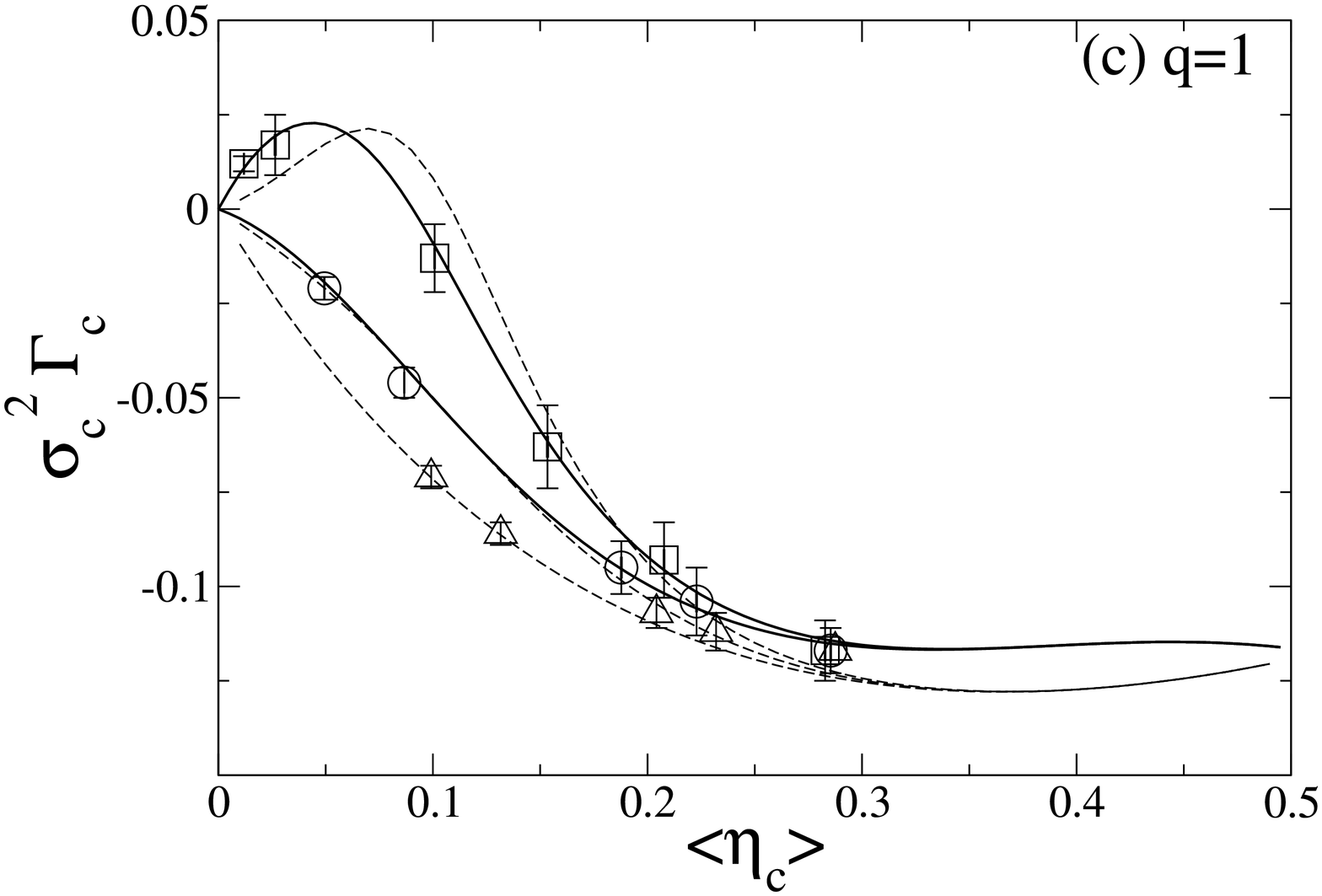}
\includegraphics[width=8cm]{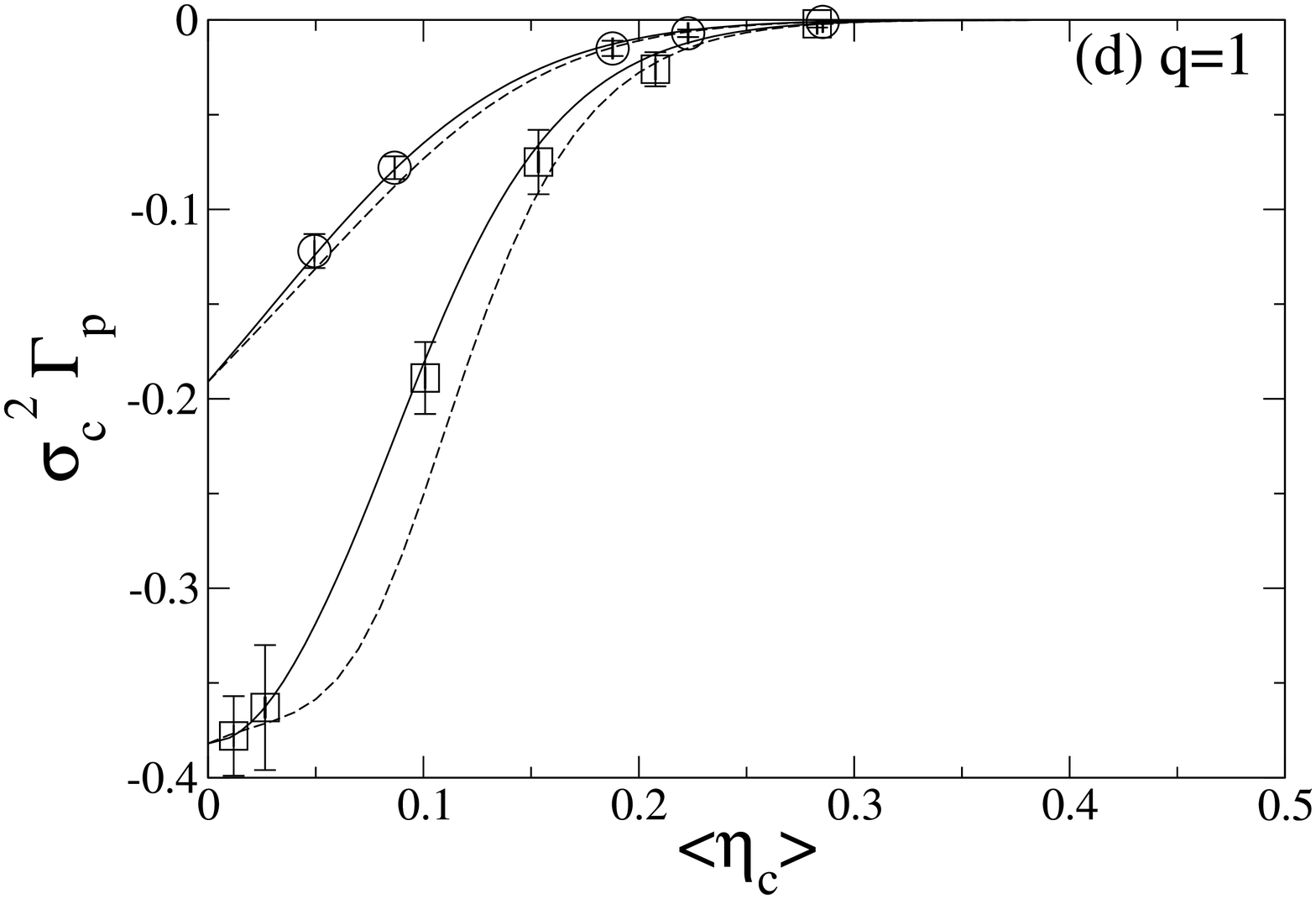}
\end{center}
\caption{The adsorption $\Gamma \sigma_c^2$ of the colloid-polymer mixture at a hard wall as a function of the average colloid packing fraction $\langle \eta_c \rangle$.
Simulation results for polymer reservoir packing fraction 
$\eta_p^r=0$ (open triangles), $\eta_p^r=0.2$ (open circles), and $\eta_p^r=0.4$ (open squares) are
compared with results from DFT (solid lines) and SPT (dashed lines).
The DFT results for $\eta_p^r=0$ are omitted for clarity. 
a) Colloid adsorption for size ratio $q$=0.6; b) Polymer adsorption
for $q$=0.6; c) Colloid adsorption for $q$=1; d) Polymer adsorption
for $q$=1.}
\label{F:ads}
\end{figure}

In this section we present results on the adsorption of colloids and polymer at a hard-wall. 
We compare the simulation results with those from DFT and SPT  Eqs.\ (\ref{Gammac}) and (\ref{Gammap}).
In Figs.\ \ref{F:ads}a and \ref{F:ads}b,  we show the results on  the colloidal
adsorption while in Figs.\ \ref{F:ads}c and  \ref{F:ads}d,  we  show the results on polymer adsorption 
for size ratio $q$=0.6 and $q$=1, respectively.  
We notice that increasing the number of polymer in the system (i.e. increasing $\eta_p^r$) the adsorption of colloids increases; the colloids are attracted at the hard wall by the depletion interaction. 
As shown by the polymer adsorption the increase in number of colloidal particle at the walls is followed by a decrease of the number of adsorbed polymer while increasing the total number of polymers in the system. 
The agreement between simulations and DFT  is good.
This is not surprising since the DFT is known to provide an accurate description of the colloid-polymer mixture at a planar hard-wall \cite{JMB03}. 
The SPT equations reproduce the $\eta_c$=0 limit correctly. 
For  $\eta_c \neq 0$ the essential features are reproduced but with low accuracy. We also
note that the differences in SPT are larger for increasing polymer
reservoir packing fraction and for  size ratio $q$=0.6. The SPT
performance is worse when the number of polymer in the mixture is
relatively high compared to the number of colloids.

\subsection{Liquid-gas interfacial tension}
\label{S:reslg}
\begin{figure}[p]
\begin{center}
\includegraphics[width=14cm]{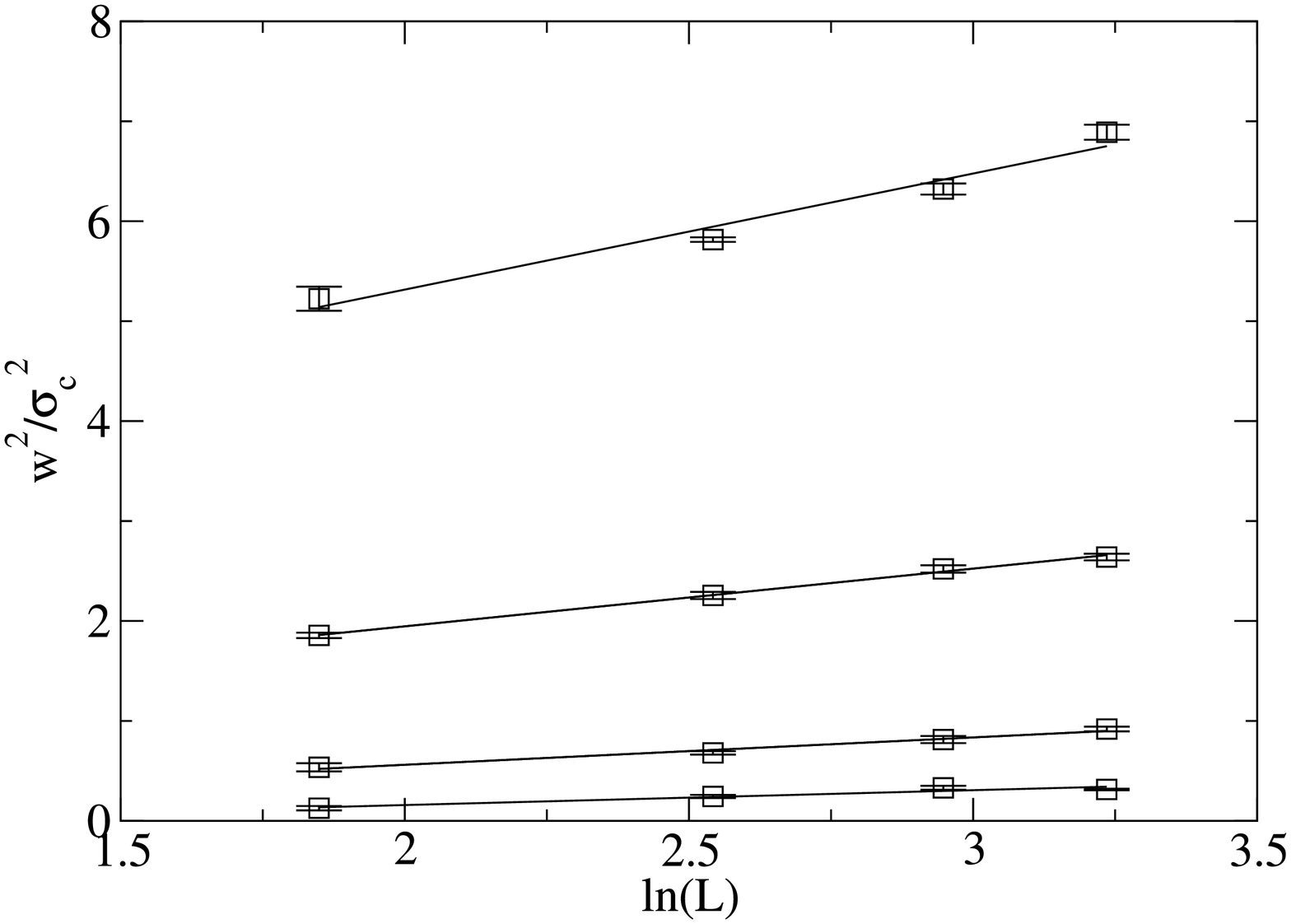}
\end{center}
\caption{The squared width of the liquid-gas interface of the model
colloid-polymer mixture with size ratio $q=1$ for polymer reservoir
packing fractions $\eta_p^r=$0.95, 1.05, 1.4, 2 (from top to bottom),
as a function of the logarithm of the lateral dimension of the
interface.  Symbols denote simulation results, lines indicate linear
fits to the data.}
\label{F:1.0fit}
\end{figure}

\begin{figure}[p]
\begin{center}
\includegraphics[width=11cm]{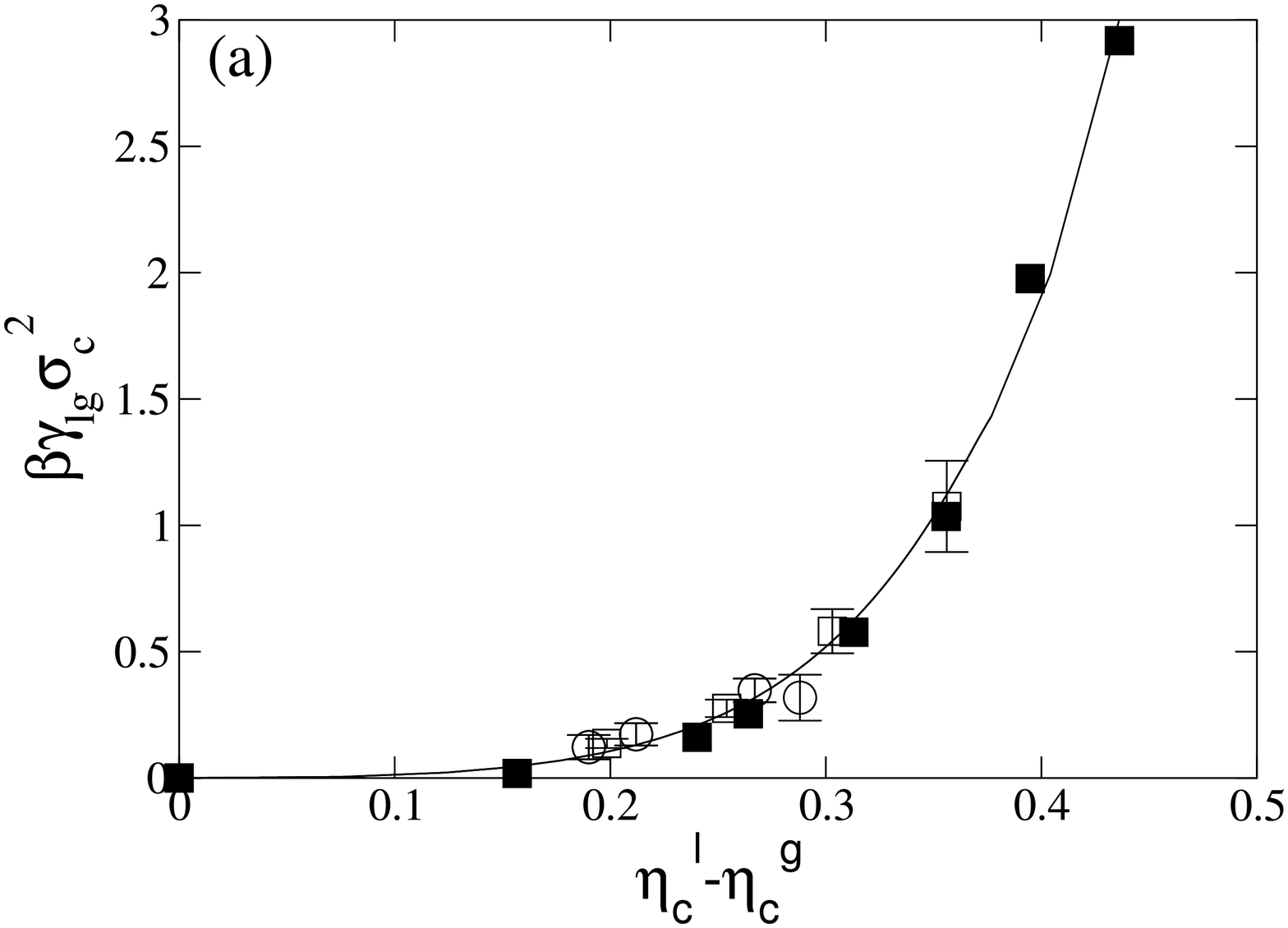}
\includegraphics[width=11cm]{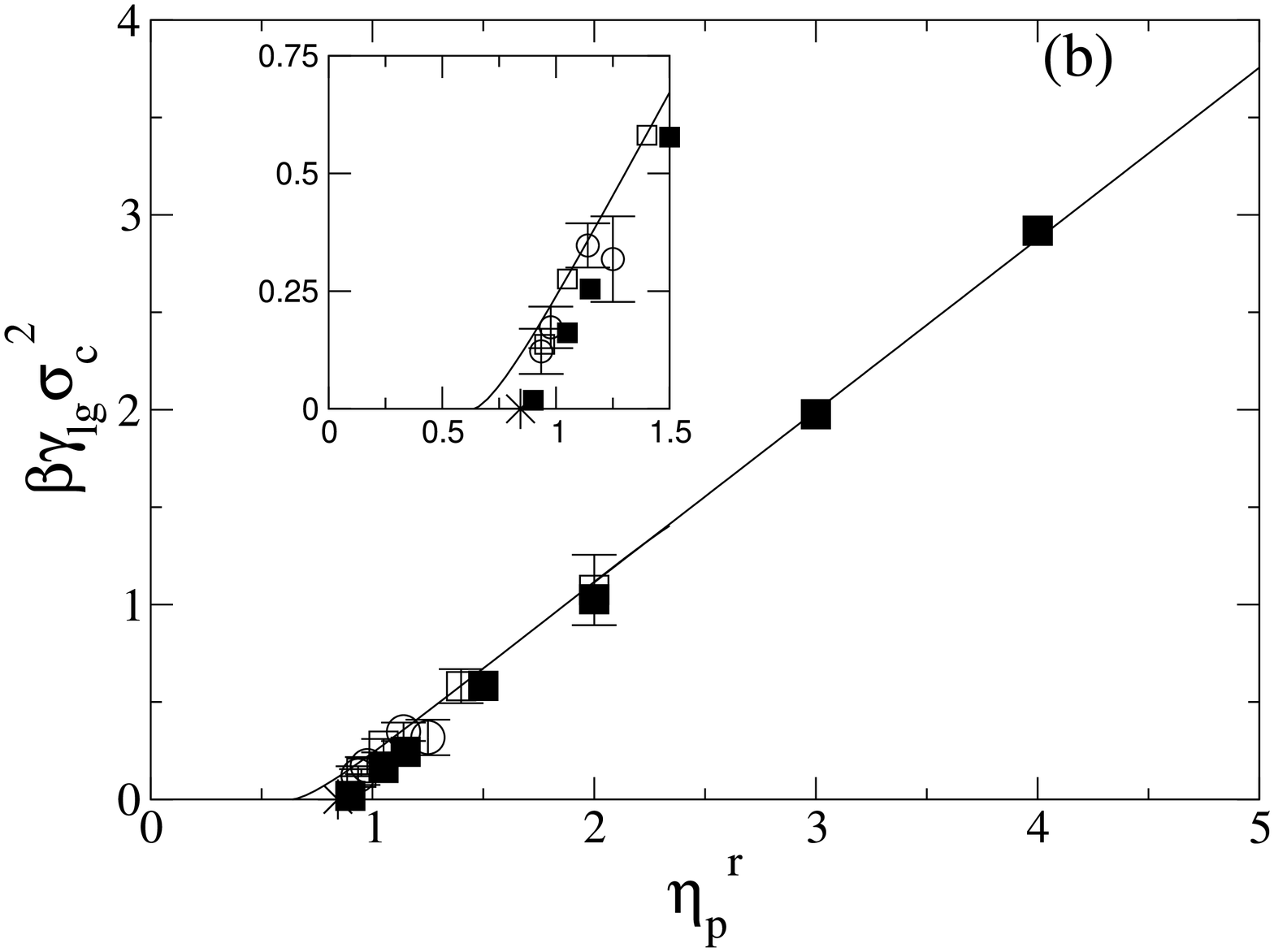}
\end{center}
\caption{The reduced liquid-gas interfacial tension $\beta \gamma
\sigma_c^2$ of the model colloid-polymer mixtures with size ratio
$q=1$.  Open squares denote simulation results from the capillary wave
method. Filled squares indicate simulation results using the
probability distribution method. Open circles refer to simulation
results employing the difference $\gamma_{\rm wg}-\gamma_{\rm wl}$. The solid curves indicates the DFT results. 
The star indicate the position of the critical point, $\eta_p^r=0.86$, obtained from simulations.
  a) As a function of the
difference in gas and liquid packing fractions $\eta_c^l-\eta_c^g$ at
coexistence.  b) As a function of the polymer reservoir packing
fraction $\eta_p^r$, at
coexistence.  The inset shows the data close to the critical
point on an enlarged scale.}
\label{F:1.0g}
\end{figure}

In this section we present results on the liquid-gas interfacial tension of AOV colloid-polymer mixtures of size ratio $q$=1, using three independent simulation techniques explained above.
First we use the scaling of the interfacial width.   
The Fig.\ \ref{snap} shows typical  snapshots of the liquid-gas interface
for $\eta_p^r=$0.95, 1.05, 1.4,  and 2.0. 
One observes clearly that the difference in densities of the two
coexisting phases increases with increasing $\eta_p^r$. 
Moreover, the liquid-gas interface becomes sharper upon increasing
$\eta_p^r$. 
The interfacial order parameter $\Psi(z)$ is determined and fitted to an error function to extract the interfacial width. 
In Fig.\ \ref{F:1.0fit} we plot the square of the
interfacial width as a function of the logarithm of the lateral
dimension $L$. Employing a linear fit to our data and using Eq.\
(\ref{sizedep}), we determine $\gamma_{\lg}$.
In Fig.\ \ref{F:1.0g}a and Fig.\ \ref{F:1.0g}b  we plot the
liquid-gas interfacial tension as a function of $\eta_c^l -
\eta_c^g$ and $\eta_p^r$ respectively. We determine the colloid
densities $\eta_c^l$ and $\eta_c^g$ at coexistence from the
measured density profile $\rho_c(z)$. 

\begin{figure}[p]
\begin{center}
\includegraphics[width=14cm]{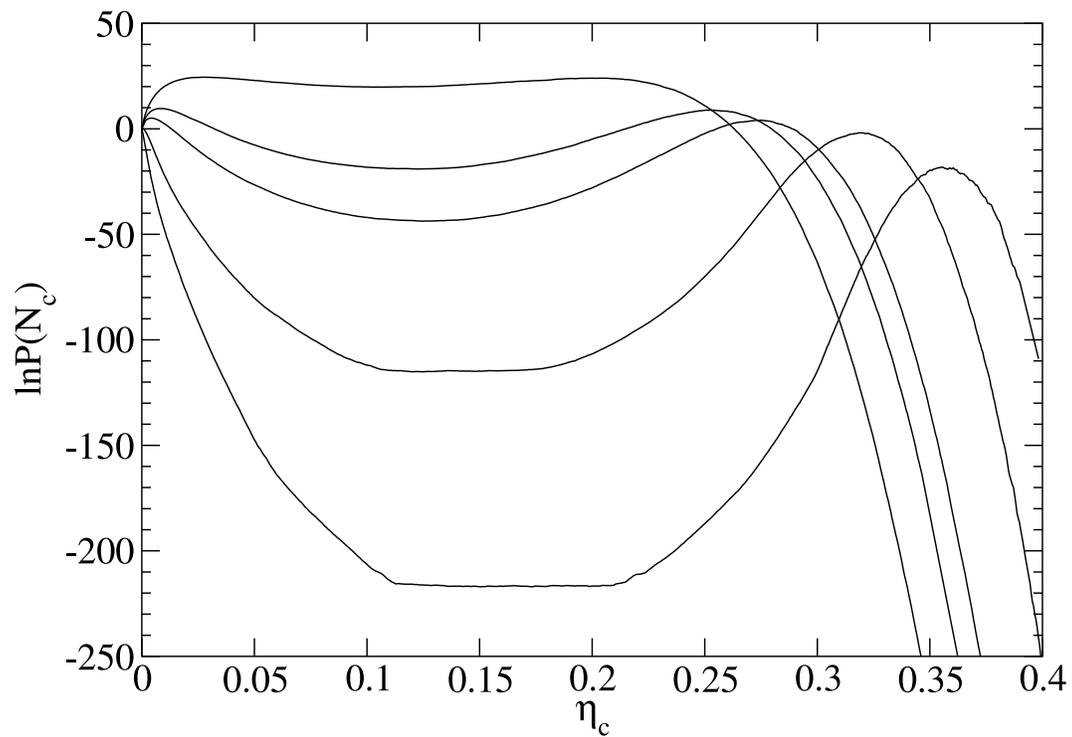}
\end{center}
\caption{Logarithm of the probability $P(N_c)$ (not normalized, as P(0) is taken to be 1) of finding
$N_c$ colloids with diameter $\sigma_c$ for a colloid-polymer mixture
with size ratio $q=1$ in a cubic box with volume $V=1000 \sigma_c^3$
at varying polymer reservoir packing fraction $\eta_p^r=$0.9, 1.05,
1.15, 1.5, and 2 (from top to bottom), as a function of $\eta_c$. All statepoints are at coexistence. }
\label{prob}
\end{figure}

\begin{figure}
\begin{center}
\includegraphics[width=14cm]{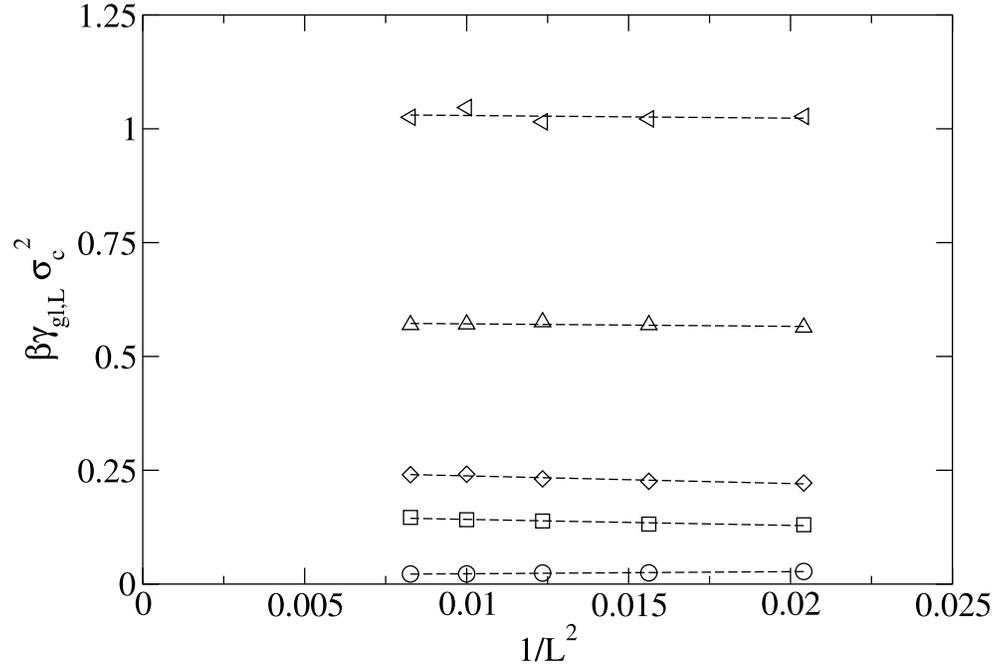}
\end{center}
\caption{The system size dependence of the (scaled) liquid-gas
interfacial tension $\beta \gamma_{\lg,L} \sigma_c^2$ of a
colloid-polymer mixture with size ratio $q=\sigma_p/\sigma_c=1$ in a
cubic box with volume $V=L^3 \sigma_c^3$ at varying polymer reservoir
packing fraction $\eta_p^r=0.9, 1.05, 1.15, 1.5,$ and 2.0 (from bottom
to top). The dashed lines are least-squares linear fits. }
\label{infinity}
\end{figure}

Than in the second method we determined the probability distributions $P(N_c)|_{z_c,\eta_p^r}$ for $\eta_p^r=$0.9, 1.05, 1.15, 1.5, 2.0, 3.0, and 4.0 in   a cubic simulation box of length $L = 7, 8, 9, 10, $ and 11. 
 We show the probability distribution for box length $L=10$ for varying $\eta_p^r$ in Fig.\ \ref{prob}. 
 We observe clearly that the  density jump at coexistence and $\gamma_{\lg,L}$, i.e., the difference of the maxima and minimum, increases with $\eta_p^r$. 
We determined $\gamma_{\lg}$ for an
infinite system by plotting $\gamma_{\lg,l}$ as a function of
$1/L^2$ in Fig.\ \ref{infinity} and extrapolating, $L\rightarrow
\infty$, according to Eq.\ (\ref{Eq:fit1}).
Again,  we plot the liquid-gas interfacial tension as a function
of $\eta_c^l - \eta_c^g$ and $\eta_p^r$ in Figs.\ \ref{F:1.0g}a
and Fig.\ \ref{F:1.0g}b,  respectively. We determined $\eta_c^g$
and $\eta_c^l$ at coexistence from \begin{equation} \eta_c^g =
\frac{ 2\pi\sigma_c^3}{6} \int_0^{\langle N_c \rangle }
N_cP(N_c)|_{z_c',\eta_p^r}dN_c \hspace{1cm}  \eta_c^l =\frac{
2\pi\sigma_c^3}{6} \int_{\langle N_c \rangle }^{\infty}
N_cP(N_c)|_{z_c',\eta_p^r}dN_c
\end{equation} where the factor 2 arises from the normalization
(\ref{norm}). 
Finally, we also employed  Young's equation to obtain an estimate
for $\gamma_{\lg}$.  Dijkstra \et \cite{MDRvR02} studied the wetting
behavior of
AOV model colloid-polymer mixtures using computer simulations. From
measuring adsorption isotherms they concluded that the colloidal
liquid phase wets the wall completely upon approaching the gas branch
of the gas-liquid bulk binodal for values of $\eta_p^r<1.05$.
In this region the contact angle vanishes and
$\gamma_{\lg}= (\gamma_{\rm wg} - \gamma_{\rm wl})$, where
$\gamma_{\rm wg}$ and $\gamma_{\rm wl}$ are wall-fluid tensions
$\gamma_{\rm wf}$ computed at liquid-gas coexistence.
We hence carried out simulations of the wall-fluid tension for
statepoints at coexistence below the wetting transition point: $\eta_p^r =
0.935$, and $0.977$ and also slightly above the wetting transition
point: $\eta_p^r=1.14$ and $1.25$.
The coexisting densities were previously determined in Gibbs ensemble
simulations
(see Fig.\ \ref{F:binodal}). 
We stress that the results obtained on the basis of Young's equation
should be taken with great care, as the wall-gas tension in our
simulations were obtained from a gas phase in contact with a planar
hard wall, which is a metastable state with respect to the equilibrium
state of a macroscopic wetting layer adsorbed at the gas-wall
interface.

However, DFT results show that the contact angle differs only less
than few a percent when the metastable gas-wall density profile is
employed instead of the stable density profile that includes the
wetting layer. In Fig.\ \ref{F:1.0g}a and Fig.\ \ref{F:1.0g}b
we plot the difference $\gamma_{\rm wg} - \gamma_{\rm wl}$ for
$\eta_p^r=0.935, 0.977, 1.14, $ and $1.25$. 
Comparing the results
obtained for $\gamma_{\lg}$ from the three different routes, we
find good agreement. 
The agreement with DFT results is good for high values of $\eta_c^l -
\eta_c^g$ and $\eta_p^r$, but deviates close to the critical point,
as might be expected. 
Similar deviations were found for $\gamma_{\lg}$, by
Vink {\em et al.} using the probability distribution method in simulation of the full mixture for size ratio $q=0.8$ \cite{RLCV04}. 

\section{Conclusions}
\label{S:end} 
In conclusion we investigated the wall-fluid tension
of the AOV model colloid-polymer mixtures of size ratio $q=0.6$
and $q=1$ using Monte Carlo computer simulations. We used a
thermodynamic integration method and a shoulder potential
approximation for the hard-core potentials to determine the free
energy of the bulk system and the inhomogeneous system. The
wall-fluid interfacial tension is the surface excess free energy
per unit area, and is in good agreement with the DFT results. 
The SPT wall-fluid interfacial tension is in overall agreement with simulations, but the comparison is worse for increasing 
polymer reservoir packing fraction. 
We also investigated the colloid and polymer adsorption of the
colloid-polymer mixture at a planar hard wall and we found good
agreement with DFT results. We derived a SPT expression for the adsorption of colloid polymer mixtures at a hard wall. 
The expression reproduce the essential features of the adsorption, but with low accuracy.  
In addition, we studied the liquid-gas interfacial tension of the AOV model colloid-polymer mixtures of
size ratio $q=1$ using i) the dependence of the interfacial
width on the logarithm of the lateral size of the simulation box
as predicted by the capillary wave theory, ii) the probability
distribution method, and iii) Young's equation in the complete
wetting regime. We find remarkably good agreement between the
different sets of results. Moreover, as we used the effective
one-component simulations in the probability distribution method,
we were able to investigate  the interfacial tension for high
$\eta_p^r$. As the bulk binodal from DFT, equivalent to that of free-volume theory\ \cite{HNWL+92}, agrees well with the
simulation results for $\eta_p^r>1.5$, one might expect a similar
agreement for the interfacial tension.  We found that our
simulation results for $\gamma_{\lg}$ approaches the DFT results
upon increasing $\eta_p^r$ and agree well with the DFT results only for
$\eta_p^r>3$. Close to the critical point, deviations are found
between the simulation and DFT results as expected due to the
shift in critical point. 

The good agreement of our simulation results for wall-fluid and
fluid-fluid interfacial tensions with those of the DFT hint at a
reliable description of the magnitude of contact angle of the
colloidal liquid-gas interface and a hard wall
\cite{PPFW04}. Nevertheless, carrying out detailed simulations at bulk
coexistence in the (numerically very demanding) partial wetting regime
of large $\eta_p^r$ is an interesting issue for future work, in
particular in light of the fact that the contact angle can be measured
in a direct (though not easy) way in experiments
\cite{DGALA04,WKW03,WKW03-a}.  
Furthermore our
thermodynamic integration technique is well suited to determine the
free energy of confined crystals and hence to predict the full phase
behavior of confined colloid-polymer mixtures; work along these lines
is in progress.
Finally we mention that interfacial properties at curved substrates have attracted recent interest\ \cite{RE03,RE05}; colloid-polymer 
mixtures are well-suited to investigate such situations\ \cite{DGALA04}.

\acknowledgments 
We thank R. Evans, R. Roth, R. Vink, J. Horbach, K. Binder, H. L\"owen, R. van Roij, D. Aarts, and H. Lekkerkerker 
for useful discussions. 
This work is part of the research program of
the {\em Stichting voor Fundamenteel Onderzoek der Materie} (FOM),
that is financially supported by the {\em Nederlandse Organisatie voor
Wetenschappelijk Onderzoek} (NWO).  We thank the Dutch National
Computer Facilities foundation for access to the SGI Origin 3800 and
SGI Altix 3700.  Support by the DFG SFB TR6 ``Physics of colloidal
dispersions in external fields'' is acknowledged.

\bibliographystyle{apsrev}
\bibliography{ref}

\begin{thebibliography}{50}
\expandafter\ifx\csname natexlab\endcsname\relax\def\natexlab#1{#1}\fi
\expandafter\ifx\csname bibnamefont\endcsname\relax
  \def\bibnamefont#1{#1}\fi
\expandafter\ifx\csname bibfnamefont\endcsname\relax
  \def\bibfnamefont#1{#1}\fi
\expandafter\ifx\csname citenamefont\endcsname\relax
  \def\citenamefont#1{#1}\fi
\expandafter\ifx\csname url\endcsname\relax
  \def\url#1{\texttt{#1}}\fi
\expandafter\ifx\csname urlprefix\endcsname\relax\def\urlprefix{URL }\fi
\providecommand{\bibinfo}[2]{#2}
\providecommand{\eprint}[2][]{\url{#2}}

\bibitem[{\citenamefont{{W. C. K. Poon}}(2002)}]{WCKP02}
\bibinfo{author}{\bibnamefont{{W. C. K. Poon}}}, \bibinfo{journal}{J. Phys.:
  Condens. Matter} \textbf{\bibinfo{volume}{14}}, \bibinfo{pages}{R859}
  (\bibinfo{year}{2002}).

\bibitem[{\citenamefont{Tuinier et~al.}(2003)\citenamefont{Tuinier, Rieger, and
  de~Kruif}}]{RT03}
\bibinfo{author}{\bibfnamefont{R.}~\bibnamefont{Tuinier}},
  \bibinfo{author}{\bibfnamefont{J.}~\bibnamefont{Rieger}}, \bibnamefont{and}
  \bibinfo{author}{\bibfnamefont{C.~G.} \bibnamefont{de~Kruif}},
  \bibinfo{journal}{Adv. Coll. Interf. Sci.} \textbf{\bibinfo{volume}{103}},
  \bibinfo{pages}{1} (\bibinfo{year}{2003}).

\bibitem[{\citenamefont{{J. M. Brader} et~al.}(2003)\citenamefont{{J. M.
  Brader}, {R. Evans}, and {M. Schmidt}}}]{JMB03}
\bibinfo{author}{\bibnamefont{{J. M. Brader}}},
  \bibinfo{author}{\bibnamefont{{R. Evans}}}, \bibnamefont{and}
  \bibinfo{author}{\bibnamefont{{M. Schmidt}}}, \bibinfo{journal}{Mol. Phys.}
  \textbf{\bibinfo{volume}{101}}, \bibinfo{pages}{3349} (\bibinfo{year}{2003}).

\bibitem[{\citenamefont{{S. Asakura} and {F. Oosawa}}(1954)}]{SAFO54}
\bibinfo{author}{\bibnamefont{{S. Asakura}}} \bibnamefont{and}
  \bibinfo{author}{\bibnamefont{{F. Oosawa}}}, \bibinfo{journal}{J. Chem.
  Phys.} \textbf{\bibinfo{volume}{22}}, \bibinfo{pages}{1255}
  (\bibinfo{year}{1954}).

\bibitem[{\citenamefont{{A. Vrij}}(1976)}]{AV76}
\bibinfo{author}{\bibnamefont{{A. Vrij}}}, \bibinfo{journal}{Pure Appl. Chem.}
  \textbf{\bibinfo{volume}{48}}, \bibinfo{pages}{471} (\bibinfo{year}{1976}).

\bibitem[{\citenamefont{{D. G. A. L. Aarts}
  et~al.}(2004{\natexlab{a}})\citenamefont{{D. G. A. L. Aarts}, {M. Schmidt},
  and {H. N. W. Lekkerkerker}}}]{DGALA04b}
\bibinfo{author}{\bibnamefont{{D. G. A. L. Aarts}}},
  \bibinfo{author}{\bibnamefont{{M. Schmidt}}}, \bibnamefont{and}
  \bibinfo{author}{\bibnamefont{{H. N. W. Lekkerkerker}}},
  \bibinfo{journal}{Science} \textbf{\bibinfo{volume}{304}},
  \bibinfo{pages}{847} (\bibinfo{year}{2004}{\natexlab{a}}).

\bibitem[{\citenamefont{{W. K. Wijting}
  et~al.}(2003{\natexlab{a}})\citenamefont{{W. K. Wijting}, {N. A. M.
  Besseling}, and {M. A. Cohen Stuart}}}]{WKW03-a}
\bibinfo{author}{\bibnamefont{{W. K. Wijting}}},
  \bibinfo{author}{\bibnamefont{{N. A. M. Besseling}}}, \bibnamefont{and}
  \bibinfo{author}{\bibnamefont{{M. A. Cohen Stuart}}}, \bibinfo{journal}{J.
  Phys. Chem. B} \textbf{\bibinfo{volume}{107}}, \bibinfo{pages}{10565}
  (\bibinfo{year}{2003}{\natexlab{a}}).

\bibitem[{\citenamefont{{D. G. A. L. Aarts} et~al.}(2003)\citenamefont{{D. G.
  A. L. Aarts}, {J. H. van der Wiel}, and {H. N. W. Lekkerkerker}}}]{DGALA03}
\bibinfo{author}{\bibnamefont{{D. G. A. L. Aarts}}},
  \bibinfo{author}{\bibnamefont{{J. H. van der Wiel}}}, \bibnamefont{and}
  \bibinfo{author}{\bibnamefont{{H. N. W. Lekkerkerker}}}, \bibinfo{journal}{J.
  Phys.: Condens. Matter} \textbf{\bibinfo{volume}{15}}, \bibinfo{pages}{S245}
  (\bibinfo{year}{2003}).

\bibitem[{\citenamefont{{W. K. Wijting}
  et~al.}(2003{\natexlab{b}})\citenamefont{{W. K. Wijting}, {N. A. M.
  Besseling}, and {M. A. Cohen Stuart}}}]{WKW03}
\bibinfo{author}{\bibnamefont{{W. K. Wijting}}},
  \bibinfo{author}{\bibnamefont{{N. A. M. Besseling}}}, \bibnamefont{and}
  \bibinfo{author}{\bibnamefont{{M. A. Cohen Stuart}}}, \bibinfo{journal}{Phys.
  Rev. Lett.} \textbf{\bibinfo{volume}{90}}, \bibinfo{pages}{196101}
  (\bibinfo{year}{2003}{\natexlab{b}}).

\bibitem[{\citenamefont{{D. G. A. L. Aarts} and {H. N. W.
  Lekkerkerker}}(2004)}]{DGALA04}
\bibinfo{author}{\bibnamefont{{D. G. A. L. Aarts}}} \bibnamefont{and}
  \bibinfo{author}{\bibnamefont{{H. N. W. Lekkerkerker}}}, \bibinfo{journal}{J.
  Phys.: Condens. Matter} \textbf{\bibinfo{volume}{16}}, \bibinfo{pages}{S4231}
  (\bibinfo{year}{2004}).

\bibitem[{\citenamefont{{H. N. W. Lekkerkerker} et~al.}(1992)\citenamefont{{H.
  N. W. Lekkerkerker}, {W. C. K. Poon}, {P. N. Pusey}, {A. Stroobants}, and {P.
  B. Warren}}}]{HNWL+92}
\bibinfo{author}{\bibnamefont{{H. N. W. Lekkerkerker}}},
  \bibinfo{author}{\bibnamefont{{W. C. K. Poon}}},
  \bibinfo{author}{\bibnamefont{{P. N. Pusey}}},
  \bibinfo{author}{\bibnamefont{{A. Stroobants}}}, \bibnamefont{and}
  \bibinfo{author}{\bibnamefont{{P. B. Warren}}}, \bibinfo{journal}{Europhys.
  Lett.} \textbf{\bibinfo{volume}{20}}, \bibinfo{pages}{559}
  (\bibinfo{year}{1992}).

\bibitem[{\citenamefont{{D. G. A. L. Aarts}
  et~al.}(2004{\natexlab{b}})\citenamefont{{D. G. A. L. Aarts}, {R. P. A.
  Dullens}, {H. N. W. Lekkerkerker}, {D. Bonn}, and {R. van Roij}}}]{DGALA+04}
\bibinfo{author}{\bibnamefont{{D. G. A. L. Aarts}}},
  \bibinfo{author}{\bibnamefont{{R. P. A. Dullens}}},
  \bibinfo{author}{\bibnamefont{{H. N. W. Lekkerkerker}}},
  \bibinfo{author}{\bibnamefont{{D. Bonn}}}, \bibnamefont{and}
  \bibinfo{author}{\bibnamefont{{R. van Roij}}}, \bibinfo{journal}{J. Chem.
  Phys.} \textbf{\bibinfo{volume}{120}}, \bibinfo{pages}{1973}
  (\bibinfo{year}{2004}{\natexlab{b}}).

\bibitem[{\citenamefont{{P. P. F. Wessels}
  et~al.}(2004{\natexlab{a}})\citenamefont{{P. P. F. Wessels}, {M. Schmidt},
  and {H. L\"owen}}}]{PPFW04}
\bibinfo{author}{\bibnamefont{{P. P. F. Wessels}}},
  \bibinfo{author}{\bibnamefont{{M. Schmidt}}}, \bibnamefont{and}
  \bibinfo{author}{\bibnamefont{{H. L\"owen}}}, \bibinfo{journal}{J. Phys.:
  Condens. Matter} \textbf{\bibinfo{volume}{16}}, \bibinfo{pages}{S4169}
  (\bibinfo{year}{2004}{\natexlab{a}}).

\bibitem[{\citenamefont{{P. P. F. Wessels}
  et~al.}(2004{\natexlab{b}})\citenamefont{{P. P. F. Wessels}, {M. Schmidt},
  and {H. L\"owen}}}]{PPFW04a}
\bibinfo{author}{\bibnamefont{{P. P. F. Wessels}}},
  \bibinfo{author}{\bibnamefont{{M. Schmidt}}}, \bibnamefont{and}
  \bibinfo{author}{\bibnamefont{{H. L\"owen}}}, \bibinfo{journal}{J. Phys.:
  Condens. Matter} \textbf{\bibinfo{volume}{16}}, \bibinfo{pages}{L1}
  (\bibinfo{year}{2004}{\natexlab{b}}).

\bibitem[{\citenamefont{Meijer and Frenkel}(1994)}]{EJM94}
\bibinfo{author}{\bibfnamefont{E.~J.} \bibnamefont{Meijer}} \bibnamefont{and}
  \bibinfo{author}{\bibfnamefont{D.}~\bibnamefont{Frenkel}},
  \bibinfo{journal}{J. Chem. Phys.} \textbf{\bibinfo{volume}{100}},
  \bibinfo{pages}{6873} (\bibinfo{year}{1994}).

\bibitem[{\citenamefont{{M. Dijkstra} et~al.}(1999)\citenamefont{{M. Dijkstra},
  {J. M. Brader}, and {R. Evans}}}]{MDJMBRE99}
\bibinfo{author}{\bibnamefont{{M. Dijkstra}}},
  \bibinfo{author}{\bibnamefont{{J. M. Brader}}}, \bibnamefont{and}
  \bibinfo{author}{\bibnamefont{{R. Evans}}}, \bibinfo{journal}{J. Phys.:
  Condens. Matter} \textbf{\bibinfo{volume}{11}}, \bibinfo{pages}{10079}
  (\bibinfo{year}{1999}).

\bibitem[{\citenamefont{{M. Dijkstra} and {R. van Roij}}(2002)}]{MDRvR02}
\bibinfo{author}{\bibnamefont{{M. Dijkstra}}} \bibnamefont{and}
  \bibinfo{author}{\bibnamefont{{R. van Roij}}}, \bibinfo{journal}{Phys. Rev.
  Lett.} \textbf{\bibinfo{volume}{89}}, \bibinfo{pages}{208303}
  (\bibinfo{year}{2002}).

\bibitem[{\citenamefont{{P. G. Bolhuis} et~al.}(2002)\citenamefont{{P. G.
  Bolhuis}, {A. A. Louis}, and {J.-P. Hansen}}}]{PBAAL02-a}
\bibinfo{author}{\bibnamefont{{P. G. Bolhuis}}},
  \bibinfo{author}{\bibnamefont{{A. A. Louis}}}, \bibnamefont{and}
  \bibinfo{author}{\bibnamefont{{J.-P. Hansen}}}, \bibinfo{journal}{Phys. Rev.
  Lett.} \textbf{\bibinfo{volume}{89}}, \bibinfo{pages}{128302}
  (\bibinfo{year}{2002}).

\bibitem[{\citenamefont{{M. Schmidt} et~al.}(2003)\citenamefont{{M. Schmidt},
  {A. Fortini}, and {M. Dijkstra}}}]{MSAF03}
\bibinfo{author}{\bibnamefont{{M. Schmidt}}}, \bibinfo{author}{\bibnamefont{{A.
  Fortini}}}, \bibnamefont{and} \bibinfo{author}{\bibnamefont{{M. Dijkstra}}},
  \bibinfo{journal}{J. Phys.: Condens. Matter} \textbf{\bibinfo{volume}{15}},
  \bibinfo{pages}{S3411} (\bibinfo{year}{2003}).

\bibitem[{\citenamefont{{M. Schmidt} et~al.}(2004)\citenamefont{{M. Schmidt},
  {A. Fortini}, and {M. Dijkstra}}}]{MSAF04}
\bibinfo{author}{\bibnamefont{{M. Schmidt}}}, \bibinfo{author}{\bibnamefont{{A.
  Fortini}}}, \bibnamefont{and} \bibinfo{author}{\bibnamefont{{M. Dijkstra}}},
  \bibinfo{journal}{J. Phys.: Condens. Matter} \textbf{\bibinfo{volume}{16}},
  \bibinfo{pages}{S4159} (\bibinfo{year}{2004}).

\bibitem[{\citenamefont{{R. L. C. Vink} and {J.
  Horbach}}(2004{\natexlab{a}})}]{RLCV03}
\bibinfo{author}{\bibnamefont{{R. L. C. Vink}}} \bibnamefont{and}
  \bibinfo{author}{\bibnamefont{{J. Horbach}}}, \bibinfo{journal}{J. Chem.
  Phys.} \textbf{\bibinfo{volume}{121}}, \bibinfo{pages}{3253}
  (\bibinfo{year}{2004}{\natexlab{a}}).

\bibitem[{\citenamefont{{R. L. C. Vink} and {J.
  Horbach}}(2004{\natexlab{b}})}]{RLCV04}
\bibinfo{author}{\bibnamefont{{R. L. C. Vink}}} \bibnamefont{and}
  \bibinfo{author}{\bibnamefont{{J. Horbach}}}, \bibinfo{journal}{J. Phys.:
  Condens. Matter} \textbf{\bibinfo{volume}{16}}, \bibinfo{pages}{S3807}
  (\bibinfo{year}{2004}{\natexlab{b}}).

\bibitem[{\citenamefont{{R. L. C. Vink}}(2004)}]{RLCV04b}
\bibinfo{author}{\bibnamefont{{R. L. C. Vink}}}, \emph{\bibinfo{title}{Entropy
  driven phase separation}}, vol.~\bibinfo{volume}{18} of
  \emph{\bibinfo{series}{Computer Simulation Studies in Condensed Matter
  Physics}} (\bibinfo{publisher}{Eds.\ {\em S.P. Lewis} and {\em H.B.
  Schuettler}, Berlin: Springer}, \bibinfo{year}{2004}).

\bibitem[{\citenamefont{{Dzubiella {\it et al.}}}(2001)}]{JDAJ+01}
\bibinfo{author}{\bibfnamefont{J.}~\bibnamefont{{Dzubiella {\it et al.}}}},
  \bibinfo{journal}{Phys. Rev. E} \textbf{\bibinfo{volume}{64}},
  \bibinfo{pages}{010401(R)} (\bibinfo{year}{2001}).

\bibitem[{\citenamefont{{M. Schmidt} et~al.}(2000)\citenamefont{{M. Schmidt},
  {H. L\"owen}, {J. M. Brader}, and {R. Evans}}}]{MSHL+00}
\bibinfo{author}{\bibnamefont{{M. Schmidt}}}, \bibinfo{author}{\bibnamefont{{H.
  L\"owen}}}, \bibinfo{author}{\bibnamefont{{J. M. Brader}}}, \bibnamefont{and}
  \bibinfo{author}{\bibnamefont{{R. Evans}}}, \bibinfo{journal}{Phys. Rev.
  Lett.} \textbf{\bibinfo{volume}{85}}, \bibinfo{pages}{1934}
  (\bibinfo{year}{2000}).

\bibitem[{\citenamefont{{Y. Rosenfeld}}(1989)}]{YR89}
\bibinfo{author}{\bibnamefont{{Y. Rosenfeld}}}, \bibinfo{journal}{Phys. Rev.
  Lett.} \textbf{\bibinfo{volume}{63}}, \bibinfo{pages}{980}
  (\bibinfo{year}{1989}).

\bibitem[{\citenamefont{{M. Heni} and {H. L\"owen}}(1999)}]{MHHL99}
\bibinfo{author}{\bibnamefont{{M. Heni}}} \bibnamefont{and}
  \bibinfo{author}{\bibnamefont{{H. L\"owen}}}, \bibinfo{journal}{Phys. Rev. E}
  \textbf{\bibinfo{volume}{60}}, \bibinfo{pages}{7057} (\bibinfo{year}{1999}).

\bibitem[{\citenamefont{{J. R. Henderson} and {F. van Swol}}(1984)}]{JRH84}
\bibinfo{author}{\bibnamefont{{J. R. Henderson}}} \bibnamefont{and}
  \bibinfo{author}{\bibnamefont{{F. van Swol}}}, \bibinfo{journal}{Mol. Phys.}
  \textbf{\bibinfo{volume}{51}}, \bibinfo{pages}{991} (\bibinfo{year}{1984}).

\bibitem[{\citenamefont{{H. Heinz} et~al.}(2004)\citenamefont{{H. Heinz}, {W.
  Paul}, and {K. Binder}}}]{HHWP04}
\bibinfo{author}{\bibnamefont{{H. Heinz}}}, \bibinfo{author}{\bibnamefont{{W.
  Paul}}}, \bibnamefont{and} \bibinfo{author}{\bibnamefont{{K. Binder}}},
  \bibinfo{journal}{cond-mat/0309014}  (\bibinfo{year}{2004}).

\bibitem[{\citenamefont{{K. Binder}}(1982)}]{KB82}
\bibinfo{author}{\bibnamefont{{K. Binder}}}, \bibinfo{journal}{Phys. Rev. A}
  \textbf{\bibinfo{volume}{25}}, \bibinfo{pages}{1699} (\bibinfo{year}{1982}).

\bibitem[{\citenamefont{{J. J. Potoff} and {A. Z.
  Panagiotopoulos}}(2000)}]{JJP00}
\bibinfo{author}{\bibnamefont{{J. J. Potoff}}} \bibnamefont{and}
  \bibinfo{author}{\bibnamefont{{A. Z. Panagiotopoulos}}}, \bibinfo{journal}{J.
  Chem. Phys.} \textbf{\bibinfo{volume}{112}}, \bibinfo{pages}{6411}
  (\bibinfo{year}{2000}).

\bibitem[{\citenamefont{{M. M\"uller} and {L. G. MacDowell}}(2003)}]{MMLGM03}
\bibinfo{author}{\bibnamefont{{M. M\"uller}}} \bibnamefont{and}
  \bibinfo{author}{\bibnamefont{{L. G. MacDowell}}}, \bibinfo{journal}{J.
  Phys.: Condens. Matter} \textbf{\bibinfo{volume}{15}}, \bibinfo{pages}{R609}
  (\bibinfo{year}{2003}).

\bibitem[{\citenamefont{Vink et~al.}(2004{\natexlab{a}})\citenamefont{Vink,
  Horbach, and Binder}}]{RLCV05}
\bibinfo{author}{\bibfnamefont{R.~L.~C.} \bibnamefont{Vink}},
  \bibinfo{author}{\bibfnamefont{J.}~\bibnamefont{Horbach}}, \bibnamefont{and}
  \bibinfo{author}{\bibfnamefont{K.}~\bibnamefont{Binder}},
  \bibinfo{journal}{cond-mat/0409099}  (\bibinfo{year}{2004}{\natexlab{a}}).

\bibitem[{\citenamefont{{S. W. Sides} et~al.}(1999)\citenamefont{{S. W. Sides},
  {G. S. Grest}, and {M.-D. Lacasse}}}]{sides}
\bibinfo{author}{\bibnamefont{{S. W. Sides}}},
  \bibinfo{author}{\bibnamefont{{G. S. Grest}}}, \bibnamefont{and}
  \bibinfo{author}{\bibnamefont{{M.-D. Lacasse}}}, \bibinfo{journal}{Phys. Rev.
  E} \textbf{\bibinfo{volume}{60}}, \bibinfo{pages}{6708}
  (\bibinfo{year}{1999}).

\bibitem[{\citenamefont{{M.-D. Lacasse} et~al.}(1998)\citenamefont{{M.-D.
  Lacasse}, {G. S. Grest}, and {A. J. Levine}}}]{lacasse}
\bibinfo{author}{\bibnamefont{{M.-D. Lacasse}}},
  \bibinfo{author}{\bibnamefont{{G. S. Grest}}}, \bibnamefont{and}
  \bibinfo{author}{\bibnamefont{{A. J. Levine}}}, \bibinfo{journal}{Phys. Rev.
  Lett.} \textbf{\bibinfo{volume}{80}}, \bibinfo{pages}{309}
  (\bibinfo{year}{1998}).

\bibitem[{\citenamefont{{R. Evans} and {U. Marini Bettolo
  Marconi}}(1987)}]{REUMBM87}
\bibinfo{author}{\bibnamefont{{R. Evans}}} \bibnamefont{and}
  \bibinfo{author}{\bibnamefont{{U. Marini Bettolo Marconi}}},
  \bibinfo{journal}{J. Chem. Phys.} \textbf{\bibinfo{volume}{86}},
  \bibinfo{pages}{7138} (\bibinfo{year}{1987}).

\bibitem[{\citenamefont{{M. Dijkstra}}(1997)}]{MD97b}
\bibinfo{author}{\bibnamefont{{M. Dijkstra}}}, \bibinfo{journal}{J. Chem.
  Phys.} \textbf{\bibinfo{volume}{107}}, \bibinfo{pages}{3277}
  (\bibinfo{year}{1997}).

\bibitem[{\citenamefont{{J. S. Rowlinson} and {B. Widom}}(2002)}]{JRBW82}
\bibinfo{author}{\bibnamefont{{J. S. Rowlinson}}} \bibnamefont{and}
  \bibinfo{author}{\bibnamefont{{B. Widom}}}, \emph{\bibinfo{title}{Molecular
  {T}heory of {C}apillarity}} (\bibinfo{publisher}{Dover},
  \bibinfo{address}{New York}, \bibinfo{year}{2002}).

\bibitem[{\citenamefont{{D. Frenkel} and {B. Smit}}(2002)}]{DFBS96}
\bibinfo{author}{\bibnamefont{{D. Frenkel}}} \bibnamefont{and}
  \bibinfo{author}{\bibnamefont{{B. Smit}}},
  \emph{\bibinfo{title}{Understanding {M}olecular {S}imulation 2nd edition}},
  vol.~\bibinfo{volume}{1} of \emph{\bibinfo{series}{Computational science
  series}} (\bibinfo{publisher}{Academic Press}, \bibinfo{year}{2002}).

\bibitem[{\citenamefont{{H. Reiss} et~al.}(1959)\citenamefont{{H. Reiss},
  Frisch, and Lebowitz}}]{HR59}
\bibinfo{author}{\bibnamefont{{H. Reiss}}},
  \bibinfo{author}{\bibfnamefont{H.~L.} \bibnamefont{Frisch}},
  \bibnamefont{and} \bibinfo{author}{\bibfnamefont{J.~L.}
  \bibnamefont{Lebowitz}}, \bibinfo{journal}{J. Chem. Phys.}
  \textbf{\bibinfo{volume}{31}}, \bibinfo{pages}{369} (\bibinfo{year}{1959}).

\bibitem[{\citenamefont{{H. Reiss} et~al.}(1960)\citenamefont{{H. Reiss}, {H.
  L. Frisch}, {E. Helfand}, and {J. L. Lebowitz}}}]{HR60}
\bibinfo{author}{\bibnamefont{{H. Reiss}}}, \bibinfo{author}{\bibnamefont{{H.
  L. Frisch}}}, \bibinfo{author}{\bibnamefont{{E. Helfand}}}, \bibnamefont{and}
  \bibinfo{author}{\bibnamefont{{J. L. Lebowitz}}}, \bibinfo{journal}{J. Chem.
  Phys.} \textbf{\bibinfo{volume}{32}}, \bibinfo{pages}{119}
  (\bibinfo{year}{1960}).

\bibitem[{\citenamefont{{R. Roth} and {S. Dietrich}}(2000)}]{RR00}
\bibinfo{author}{\bibnamefont{{R. Roth}}} \bibnamefont{and}
  \bibinfo{author}{\bibnamefont{{S. Dietrich}}}, \bibinfo{journal}{Phys. Rev.
  E} \textbf{\bibinfo{volume}{62}}, \bibinfo{pages}{6926}
  (\bibinfo{year}{2000}).

\bibitem[{\citenamefont{Vink et~al.}(2004{\natexlab{b}})\citenamefont{Vink,
  Horbach, and Binder}}]{RLCV05a}
\bibinfo{author}{\bibfnamefont{R.~L.~C.} \bibnamefont{Vink}},
  \bibinfo{author}{\bibfnamefont{J.}~\bibnamefont{Horbach}}, \bibnamefont{and}
  \bibinfo{author}{\bibfnamefont{K.}~\bibnamefont{Binder}},
  \bibinfo{journal}{cond-mat/0411722}  (\bibinfo{year}{2004}{\natexlab{b}}).

\bibitem[{\citenamefont{{S. Senapati} and {M. L. Berkowitz}}(2001)}]{senapati}
\bibinfo{author}{\bibnamefont{{S. Senapati}}} \bibnamefont{and}
  \bibinfo{author}{\bibnamefont{{M. L. Berkowitz}}}, \bibinfo{journal}{Phys.
  Rev. Lett.} \textbf{\bibinfo{volume}{87}}, \bibinfo{pages}{176101}
  (\bibinfo{year}{2001}).

\bibitem[{\citenamefont{{A. Aguado} et~al.}(2001)\citenamefont{{A. Aguado}, {W.
  Scott}, and {P. A. Madden}}}]{scott}
\bibinfo{author}{\bibnamefont{{A. Aguado}}}, \bibinfo{author}{\bibnamefont{{W.
  Scott}}}, \bibnamefont{and} \bibinfo{author}{\bibnamefont{{P. A. Madden}}},
  \bibinfo{journal}{J. Chem. Phys.} \textbf{\bibinfo{volume}{115}},
  \bibinfo{pages}{8612} (\bibinfo{year}{2001}).

\bibitem[{\citenamefont{Dijkstra et~al.}()\citenamefont{Dijkstra, van Roij,
  Roth, and Fortini}}]{MD05}
\bibinfo{author}{\bibfnamefont{M.}~\bibnamefont{Dijkstra}},
  \bibinfo{author}{\bibfnamefont{R.}~\bibnamefont{van Roij}},
  \bibinfo{author}{\bibfnamefont{R.}~\bibnamefont{Roth}}, \bibnamefont{and}
  \bibinfo{author}{\bibfnamefont{A.}~\bibnamefont{Fortini}}, \bibinfo{note}{to
  be published.}

\bibitem[{\citenamefont{{P. Virnau} and {M. M\"uller}}(2004)}]{virnau}
\bibinfo{author}{\bibnamefont{{P. Virnau}}} \bibnamefont{and}
  \bibinfo{author}{\bibnamefont{{M. M\"uller}}}, \bibinfo{journal}{J. Chem.
  Phys.} \textbf{\bibinfo{volume}{120}}, \bibinfo{pages}{10925}
  (\bibinfo{year}{2004}).

\bibitem[{\citenamefont{Roth and Evans}(2004)}]{RR1}
\bibinfo{author}{\bibfnamefont{R.}~\bibnamefont{Roth}} \bibnamefont{and}
  \bibinfo{author}{\bibfnamefont{R.}~\bibnamefont{Evans}},
  \bibinfo{journal}{private communication}  (\bibinfo{year}{2004}).

\bibitem[{\citenamefont{Evans et~al.}(2003)\citenamefont{Evans, R.Roth, and
  Bryk}}]{RE03}
\bibinfo{author}{\bibfnamefont{R.}~\bibnamefont{Evans}},
  \bibinfo{author}{\bibnamefont{R.Roth}}, \bibnamefont{and}
  \bibinfo{author}{\bibfnamefont{P.}~\bibnamefont{Bryk}},
  \bibinfo{journal}{Europhys. Lett.} \textbf{\bibinfo{volume}{62}},
  \bibinfo{pages}{815} (\bibinfo{year}{2003}).

\bibitem[{\citenamefont{Evans et~al.}(2004)\citenamefont{Evans, Henderson, and
  Roth}}]{RE05}
\bibinfo{author}{\bibfnamefont{R.}~\bibnamefont{Evans}},
  \bibinfo{author}{\bibfnamefont{J.}~\bibnamefont{Henderson}},
  \bibnamefont{and} \bibinfo{author}{\bibfnamefont{R.}~\bibnamefont{Roth}},
  \bibinfo{journal}{cond-mat/0410179}  (\bibinfo{year}{2004}).

\end{thebibliography}

\clearpage

\clearpage

\clearpage

\clearpage

\clearpage

\clearpage

\clearpage

\clearpage

\clearpage

\clearpage

\end{document}